\documentstyle[amssymb,oldlfont,epsf]{article}

\input psfig

\newcommand{\BY}{{\Bbb Y}}
\newcommand{\BC}{{\Bbb C}}
\newcommand{\BR}{{\Bbb R}}
\newcommand{\BZ}{{\Bbb Z}}
\newcommand{\Span}{\mathop{\rm span}\nolimits}
\newcommand{\Gr}{\mathop{\rm Gr}\nolimits}
\newcommand{\ord}{\mathop{\rm ord}\nolimits}
\newcommand{\ad}{\mathop{\rm ad}\nolimits}
\newcommand{\tr}{\mathop{\rm tr}\nolimits}
\newcommand{\diag}{\mathop{\rm diag}\nolimits}

\def\E{e}\def\I{i}
\newtheorem{definition}{Definition}
\newtheorem{theorem}{Theorem}
\newtheorem{lemma}{Lemma}
\newtheorem{proposition}{Proposition}
\newtheorem{corollary}{Corollary}
\newtheorem{remark}{Remark}
\newtheorem{claim}{Claim}
\def\proof{\medskip\par\noindent{\it Proof:\ }}
\def\endproof{{\unskip\nobreak\penalty200\hskip1.5mm\null\nobreak\hfil
\rule{1.7mm}{2.5mm}\parfillskip=0pt\finalhyphendemerits0\medskip\par}}
\begin{document}

\title{A Matrix Integral Solution to $[P,Q]=P$ and\\Matrix Laplace Transforms}
\author{M. Adler\thanks{Department of Mathematics, Brandeis University,
	Waltham, MA 02254, USA.\hfill\break
	E-mail: adler@binah.cc.brandeis.edu.
	The support of a National Science Foundation
	grant \#DMS-95-4-51179 is gratefully acknowledged.},
	A. Morozov\thanks{ITEP, Moscow, Russia. E-mail: morozov@vitep5.itep.ru.
	The hospitality of the Volterra
	Center at Brandeis University is gratefully acknowledged.},
	T. Shiota\thanks{
	Department of Mathematics, Kyoto University, Kyoto 606--01, Japan.
	E-mail: shiota@kusm.kyoto-u.ac.jp.
	The hospitality of the University
	of Louvain and Brandeis University is gratefully acknowledged.},
	P. van Moerbeke\thanks{
	Department of Mathematics, Brandeis University,
	Waltham, MA 02254, USA and
	Department of Mathematics, Universite de Louvain,
	1348 Louvain-la-Neuve, Belgium. E-mail:
	vanmoerbeke@binah.cc.brandeis.edu,
	vanm@geom.ucl.ac.be.
	The support of
	a National Science Foundation grant \#DMS-95-4-51179, a Nato, an FNRS
	and a Francqui Foundation grant is gratefully acknowledged.}}
\date{}
\maketitle

\subsection*{Abstract}
In this paper we solve the following problems: (i) find two differential
operators $P$ and $Q$ satisfying $[P,Q]=P$, where $P$ flows according to
the KP hierarchy $\partial P/\partial t_n = [(P^{n/p})_+,P]$, with
$p := \ord P\ge2$; (ii) find a matrix integral representation for the
associated $\tau$-function.  First we construct an infinite dimensional
space ${\cal W}=\Span_\BC\bigl\{\psi_0(z),\psi_1(z),\dots\bigr\}$ of functions
of $z\in\BC$ invariant under the action of two operators, multiplication by
$z^p$ and $A_c:= z\,\partial/\partial z - z + c$.  This requirement is
satisfied, for arbitrary $p$, if $\psi_0$ is a certain function generalizing
the classical H\"ankel function (for $p=2$); our representation of the
generalized H\"ankel function as a double Laplace transform of a simple
function, which was unknown even for the $p=2$ case, enables us to represent
the $\tau$-function associated with the KP time evolution of the space
$\cal W$ as a ``double matrix Laplace transform'' in two different ways.
One representation involves an integration over the space of matrices whose
spectrum belongs to a wedge-shaped contour
$\gamma := \gamma^+ + \gamma^-\subset\BC$ defined by
$\gamma^\pm=\BR_+\E^{\pm\pi\I/p}$.  The new integrals above relate to the
matrix Laplace transforms, in contrast with the matrix Fourier transforms,
which generalize the Kontsevich integrals and solve the operator equation
$[P,Q]=1$.
\section*{Introduction}
It is a long-standing puzzle in the theory of $2d$-gravity
to find an adequate description of gravitational coupling
of $(p,q)$ minimal models. One part of it is to
find two differential operators $P$ and $Q$ of order $p$ and $q$
respectively, such that $[P,Q]=f(P)$ for some function $f$.
In the simplest case of $q=1$ and $f\equiv1$,
such description is provided by 1-matrix models, especially by the
Kontsevich integral and their generalizations; see \cite{avm,kon,UFN}.
Going along the chain, $2d$-gravity $\rightarrow$ equilateral
triangles $\rightarrow$ discrete
matrix models $\rightarrow$ Kontsevich models,
this approach has lead to the discovery of integrable structures
for non-perturbative partition functions, which
take the form of $\tau$-functions of the
KP hierarchy (see \cite{D,UFN,vM} for review and references). While similar
results are believed to be true in the general $(p,q)$-case,
the Kontsevich integral counterparts are still unknown. Note that
a minor modification of the generalized
Kontsevich integral can be interpreted as a duality transformation
between $(p,q)$ and $(q,p)$-models \cite{KM}.

So far the most promising approach for finding integrable structures
in the general $(p,q)$-case seems to be the one initiated by Kac--Schwarz
in the case $q=1$ and $f=1$.
So, the general problem comes in two stages: (1)
Find a point in Sato's Grassmannian invariant under two symmetry
operators, satisfying some commutation relation;
the existence of such a plane leads to a system of differential equations
specifying the wave function $\Psi$ and
thus to an algebra of constraints for the
$\tau$-function. (2) Find a matrix integral representation for this
$\tau$-function. Note a matrix representation,
 beyond the case $q=1$ and $f=1$, if it exists at all, was unknown.

The purpose of this paper is to find a $\tau$-function
and a matrix integral
representation for the equation $[P,Q]=P$ for $q=1$ and arbitrary $p$ .
Remarkably, the matrix integral representation can still be found,
but it is far less straightforward and considerably more involved,
than the ordinary Kontsevich integral.

The message is the following: whereas the case $[P,Q]=1$ is described by
general {\em matrix Fourier transforms}, a solution to $[P,Q]=P$ is related
to {\em double Laplace transforms}. While it is not known whether
this solution has immediate physical relevance, it may help to shed
some light on the $(p,q)$-case and on the matrix representations of
the corresponding $\tau$-functions. In particular, what are the proper
multimatrix generalizations of the Kontsevich integrals?

Note this problem has come up in the physical
literature, in various different contexts: unitary matrix models
have been written down, leading to equations $[P,Q]=P$ for differential
operators $P$ and $Q$ in
the double scaling limit; see the studies of
Dalley, Johnson, Periwal, Minahan,
Morris, Shevitz, and W\"atterstam \cite{djmw,djm,ps,ps1,m1,m2}).
In the mathematical context
(inverse scattering and monodromy preserving transformations),
see Ablowitz, Flaschka, Fokas and Newell \cite{f-a,f,fn}).
The solution provided in our paper is new and
does {\em not\/} require any scaling limit.

Consider the problem of finding a differential operator $P$
of order $p$ and another differential operator $Q$ satisfying
\begin{equation}
	[P,Q]=f(P)\,,\quad\mbox{with}\ 0\ne f(z)\in\BC[z]\,.
	\label{0.1}
\end{equation}
When $P$ is (formally) deformed with respect to the KP flows, i.e.,
$\partial P/\partial t_n=[(P^{n/p})_+,P]$, one can introduce the
corresponding deformation of $Q$ which preserves Eq.~(\ref{0.1}).
Hence (\ref{0.1}) can be considered as a condition on a solution of
the $p$-reduced KP hierarchy.

The basic ingredients of this construction are\footnote{
	$\BC[[x]]:=\{\sum_{n=0}^\infty a_nx^n\mid a_n\in\BC\}$
	is the ring of formal power series in $x$, and
	$\BC((x)):=\{\sum_{-\infty\ll n<\infty}a_nx^n\mid a_n\in\BC\}$
	is the ring of formal Laurent series in $x$.}
\begin{itemize}
\item[$\circ$]	$\psi_0\in1+z^{-1}\BC[[z^{-1}]]$,
\item[$\circ$]	$A\colon \BC((z^{-1}))\to\BC((z^{-1}))$ which increases the
	order of an element of $\BC((z^{-1}))$ in $z$ exactly by one,
\end{itemize}
so that ${\cal W}:=\Span_\BC\{\psi_0,A\psi_0,A^2\psi_0,\ldots\}$ belongs to
the big stratum of the Sato Grassmannian and satisfies
$A{\cal W}\subset{\cal W}$, such that
\begin{itemize}
\item[$\circ$]	$\psi_0$ satisfies the differential equation
	$v(z)\psi_0=F(A)\psi_0$
	for some $v(z)\in\BC((z^{-1}))$ and $F(Z)\in\BC[Z]$, so that
	$v(z){\cal W}\subset{\cal W}$ also holds.
\end{itemize}
Let $\Psi$ be the KP wave function corresponding to ${\cal W}$.
The above conditions lead to the existence of differential operators $Q$
and $P$ in $x$ such that $Q\Psi=A\Psi$ and $P\Psi=v(z)\Psi$.
If $A$ coincides with $\partial/\partial v=(1/v')\partial/\partial z$
up to the conjugation by a
function, then we have $[P,Q]=1$. And if $\psi_0$ is defined by a Fourier
transform and the action of $A$ on it can be expressed in a suitable way,
then the corresponding Hermitian {\em matrix\/} Fourier transform, properly
normalized, is the corresponding $\tau$-function.
See Sect.~\ref{Kontsevich} for details.

The matrix integral approach to (\ref{0.1}) has so far needed $\ord Q=1$ at
the initial point of the formal KP time flows, requiring $\deg_z f(z)\le1$.
The degree 0 case can be reduced to $[P,Q]=1$.
In this paper, we provide a solution to the degree 1 case, or the next
simplest instance of (\ref{0.1}), which can clearly be reduced to
\begin{equation}
	[P,Q]=P\,,
	\label{0.2}
\end{equation}
with differential operators $P$ and $Q$.
As in the case of $[P,Q]=1$, we write the $\tau$-function of its formal
KP deformation explicitly in terms of a matrix integral.

\begin{definition}
Let $-1<c<0$, $p\in\BZ$, $p\ge2$. Let ${\cal W}$ be the linear span
$$
	{\cal W}=\Span_\BC\bigl\{\psi_0(z),\psi_1(z),\psi_2(z),\ldots\bigr\}\,,
$$
of generalized H\"ankel functions,
\begin{equation}
	\psi_k(z)=\frac{p^{c+1}}{\Gamma(-c)}\int_1^\infty
	\frac{z^{-c}(uz)^k\E^{-(u-1)z}}{(u^p-1)^{c+1}}\,du\,,
	\quad k=0,1,2,\dots,
	\label{0.3}
\end{equation}
also representable as double Laplace transforms
\begin{equation}
\psi_k(z)=\frac{p^{c+1}}{2\pi \I}z^{(p-1)(c+1)}\E^z\int_0^\infty dx\,x^c
\E^{-xz^p}\int_0^\infty dy\,f_k(y)\E^{-xy^p}
\end{equation}
of the functions
\begin{equation}
	f_k(y)=(\zeta^{k+1}\E^{-\zeta y}
	-\zeta^{-k-1}\E^{-\zeta^{-1} y})y^k,
	\quad
	\mbox{$k=0,1,2,\dots,$ \ with \ $\zeta:= \E^{\pi \I/p}$.}
	\label{f_k}
\end{equation}
\end{definition}
Using the asymptotic expansion
$\psi_k(z)=z^k\bigl(1+O(1/z)\bigr)\in\BC((z^{-1}))$ as $\Re z\to\infty$,
${\cal W}$ defines a point of the Sato Grassmannian $\Gr$.
Let $\Psi$ and $\tau$ be the wave
(formal Baker--Akhiezer) function and $\tau$-function, respectively,
associated with the KP time evolution ${\cal W}^t=\E^{-\sum t_iz^i}\cal W$;
see Sects.~\ref{sect1} and \ref{sect2}.
Then we have

\begin{theorem}\label{T0}
\begin{equation}
	\Psi(x,0,z)=\E^{xz}\psi_0\bigl((1-x)z\bigr)\,,
\label{similar}
\end{equation}
and it satisfies
\begin{equation}
	\biggl(L\biggl(x-1,{\partial\over\partial x}\biggr)-z^p\biggr)
	\Psi(x,0,z)=0
	\,\ \hbox{and}\ \,
	\biggl(L\biggl(z,{\partial\over\partial z}-1\biggr)-(x-1)^p\biggr)
	\Psi(x,0,z)=0\,,
\label{1.3a}
\end{equation}
where $L(z,\partial/\partial z)$ is the monic differential operator
$$
	L\left(z,{\partial\over\partial z}\right):={1\over z^p}\Biggl(
		\prod_{i=0}^{p-1}
		\biggl(z{\partial\over\partial z}+c-i\biggr)
		-cp\prod_{i=0}^{p-2}
		\biggl(z{\partial\over\partial z}+c-i\biggr)
	\Biggr)
	=\biggl({\partial\over\partial z}\biggr)^p + \cdots\,.
$$
\end{theorem}
Note that for $p=2$,
$L(z,\partial/\partial z)=(\partial/\partial z)^2 -(c^2+c)/z^2$.

\begin{theorem}\label{T1}
Let ${\cal H}_N$ be the space of $N\times N$ Hermitian matrices, and
${\cal H}^+_N$ the subspace of ${\cal H}_N$ of positive definite
Hermitian matrices.
The corresponding $\tau$-function evaluated at
$$
	t_n:=-\frac{1}{n}\tr Z^{-n}\,,\quad
	\mbox{for $n=1,2,\dots,$ and with an $N\times N$ diagonal $Z$},
$$
is given by the following (normalized) double {\em matrix\/}
Laplace transform:
$$
	\tau(t)=
	S_1(t)\frac{\int_{{\cal H}^+_N} dX\det X^c \E^{-\tr Z^pX}
	\int_{{\cal H}^+_N}dY S_0(y)\E^{-\tr XY^p}}
	{\int_{{\cal H}_N}dX\exp\tr\left(-\frac{(X+Z)^{p+1}}{p+1}\right)_2}\,,
$$
where $(\,)_2$ denotes the terms quadratic in $X$,
$$
	S_0(y):=
	{\Delta(y^p)\over\Delta(y)^2}
	\det\bigl(f_{k-1}(y_i)\bigr)_{1\le i,k\le N}
	\quad\hbox{and}\quad
	S_1(t):=\det(Z^{(p-1)(c+1/2)})\E^{\tr Z},
$$
where $y=(y_1,\ldots,y_N)$ are the eigenvalues of $Y$,
$y^p=(y_1^p,\ldots,y_N^p)$, and
$\Delta(y):=\prod_{i>j}(y_i-y_j)=\det(y_i^{j-1})_{i,j}$, and $f_{k-1}$ are
as in (\ref{f_k}).

The function $\tau(t)$ also has the following matrix integral
representation
$$
	\tau(t)=\frac{
		\int_{{\cal H}^\gamma_N}m(dW) \int_{{\cal H}^+_N}dX
		\det X^c
		\bigl(\Delta(w^p)/\Delta(w)\bigr)
		\E^{\tr(Z-W)}\E^{\tr X(W^p-Z^p)}
	}{	\int_{{\cal H}_N}dX
		\exp\tr\left(-\frac{(X+Z)^{p+1}}{p+1}\right)_2
	}\,,
$$
integrated over the space of matrices
$$
	{\cal H}^\gamma_N=\bigl\{W =UD_\gamma U^{-1}\bigm| U\in{\bf U}(N),
	D_\gamma:=\diag(w_1,\dots,w_N)\in (\gamma)^N\bigr\},
$$
where $\gamma$
denotes a wedge-shaped contour in $\BC$, defined in Sect.~\ref{sect4}
(see Fig.~\ref{F1}),
in terms of a
complex-valued measure
$$
	m(dW)=dU\,dw\prod_{1\le i<j\le N}(w_i-w_j)^2.
$$
\end{theorem}
\begin{theorem}\label{T2}
(i)
The algebra of stabilizers of ${\cal W}$,
$$
	S_{\cal W}:= \bigl\{
	\phi(z,\partial/\partial z)\in\BC((z^{-1}))[\partial/\partial z]
	\hbox{\ such that\ }
	\phi{\cal W}\subset{\cal W}\bigr\},
$$
is generated by $A_c :=z\frac{\partial}{\partial z}-z+c$, $z^p$ and
$\xi:=z^{-p}F(A_c)$,
where $F(u)=\prod^{p-1}_0(u-i)-cp\prod^{p-2}_0(u-i)$:
$$
	S_{\cal W}=\BC[A_c,z^p,\xi]\subset\BC((z^{-1}))[\partial/\partial z].
$$
Moreover, ${\cal W}=\BC[A_c]\psi_0$, and $\psi_0$ satisfies the
differential equation
\begin{equation}
	F(A_c)\psi_0=(-z)^p\psi_0(z)\,.
	\label{1.3}
\end{equation}
\\
(ii) A family of solutions to the operator equation
$[P,Q]=P$ is given by the differential operators $P$ and $Q$,
defined equivalently by
\begin{equation}
	P\Psi=z^p\Psi\,,\quad Q\Psi={1\over p}A_c\Psi\,,
\label{eigenProblem}
\end{equation}
or by
$$
	P=S\left({d\over dx}\right)^pS^{-1}\quad\mbox{and}\quad
	Q=\frac{1}{p} (MP^{1/p}-P^{1/p}+c)\,,
$$
where $M=S\left(\sum_1^\infty k\bar t_k(d/dx)^{k-1}\right)S^{-1}$,
$\bar t_k=t_k+\delta_{k,1}x$, with wave operator
$$
S=\frac{\tau(\bar t-[(d/dx)^{-1}])}{\tau(\bar t)}\,.
$$
(iii) The function $\tau(t)$ satisfies, in terms of
the $W$-generators in Eq.~(\ref{Wgen}), the following constraints
\begin{equation}
	\sum_{\scriptstyle 0\le i\le m\atop\scriptstyle 0\le j\le i}
	\alpha_{m,i}{i\choose j}{(-1)^{i-j}\over j+1}
	W^{(j+1)}_{i+np-j}\tau(t)=a_{m,n,c}\tau(t)\,,\quad m,n=0,1,2,\ldots,
	\label{Wconstraint}
\end{equation}
for some constants $a_{m,n,c}$, where the constants
$\alpha_{n,i}$ are defined by the formula\/\footnote{
	More explicitly,
	$\alpha_{n,i}
	={1\over i!}\sum_{j=0}^i{i\choose j}(-1)^{i-j}j^n$.
	Note that it vanishes if $n>0$ and $i=0$.}
$(x\cdot d/dx)^n=\sum^n_{i=0}\alpha_{n,i}x^i (d/dx)^i$.
In particular, setting $m=1$, $\tau(t)$ satisfies Virasoro constraints
of the form (with $W_{np}^{(2)} = \sum_{i+j=np}\mathopen:J_i^{(1)}J_j^{(1)}
\mathclose:$)
\begin{equation}
	\biggl({1\over2}W_{np}^{(2)}-{\partial\over\partial t_{np+1}}-a_{1,n,c}
	\biggr)\tau = 0\,,\quad n=0,1,2,\ldots.
	\label{Virasoro}
\end{equation}
\end{theorem}

\begin{remark}\label{rmk1}
The constants $a_{m,n,c}$ in (\ref{Wconstraint}) can all be calculated;
in particular, the Virasoro constraint (\ref{Virasoro}) for $n=0$ becomes:
$$
	\left(
	\sum^\infty_1 it_i\frac{\partial}{\partial t_i}
	-\frac{\partial}{\partial t_1}-\frac{c(1+c)(p-1)}{2}
	\right)\tau=0\,.
$$
\end{remark}

\tableofcontents

\section{The KP Hierarchy}
\label{sect1}

Throughout, $x$ is a formal scalar variable near 0, and $z$ is a formal
scalar variable near $\infty$.  If $g(z)=cz^q\bigl(1+O(z^{-1})\bigr)$,
$c\ne0$, then $\ord_zg(z):=q$ is the {\em order\/} of $g(z)$.

Throughout, we denote $\partial/\partial x$ by $D$.
The algebra of ordinary pseudodifferential operators in $x$ is denoted by
${\cal D}$ (the word ``in $x$" may be dropped if there is no fear of
confusion), with its splitting ${\cal D}={\cal D}_++{\cal D}_-$
into the subalgebras of ordinary differential operators and of ordinary
pseudodifferential operators of negative order:
$$
	{\cal D}=\Biggl\{\sum_{-\infty<i\le n} a_i D^i\Biggm|
	n\in\BZ\ \mbox{arbitrary},\ a_i = a_i(x)\Biggr\}\,,
$$
$$
	A=\sum a_iD^i\in{\cal D}\quad\Rightarrow\quad
	A_+=\sum_{i\ge 0} a_i D^i\in{\cal D}_+\ \ \mbox{and}\ \
	A_-=A-A_+\in{\cal D}_-\,.
$$
The ring ${\cal D}$ acts on the space of functions of the form
$\sum_{-\infty<i\ll\infty}a_i(x)z^i\E^{xz}$ simply by extending the
formulas $D^n\E^{xz}=z^n\E^{xz}$ and $A(B\E^{xz})=(A\circ B)\E^{xz}$,
$A$,~$B\in{\cal D}$.  When $A\in{\cal D}_+$, this definition of $A(B\E^{xz})$
coincides with the usual action of $A$, as a differential operator,
on $B\E^{xz}$ as a formal series in $x$ with $z$-dependent coefficients.

A pseudodifferential operator in $x$ may depend on the KP time variables
$t=(t_1,t_2,\ldots)$ introduced below, but not on $z$ unless otherwise
noted.  We are not specific about the regularity of the coefficients of
pseudodifferential operators.  The operators $S$, $L$, $M$ etc., associated
to a point ${\cal W}$ of the big stratum $\Gr^0$ of the Sato Grassmannian
(see below) have regular (i.e., formal power series) coefficients;
otherwise, the singularities of those operators can be controled
by the Schubert stratum to which ${\cal W}\in\Gr$ belongs. In particular,
there exist $n$,~$m\ge0$ such that $x^nS$ and $S^{-1}x^m$ at $t=0$
have regular coefficients.  See \cite{Sa1} for details.

As in \cite{asvm}, we set $\bar t=(x+t_1,t_2,t_3,\ldots)$, and
$$
	\tilde\partial=\left({\partial\over\partial t_1},
	{1\over2}{\partial\over\partial t_2},
	{1\over3}{\partial\over\partial t_3},\ldots\right).
$$
The elementary Schur functions $p_n$ are defined by
$\exp\left(\sum_1^\infty t_nz^n\right)=\sum_0^\infty p_n(t)z^n$.

\subsection{KP hierarchy}

The operator $L = L(t) = D+\sum_{j=-\infty}^{-1}a_j(x,t)D^j \in{\cal D}$,
with $t=(t_1,t_2,\ldots)$, subjected to the KP equations
$$
	\frac{\partial L}{\partial t_n}=[(L^n)_+,L]\,,\quad n=1,2,\dots,
$$
is known to have the following representation in terms of
an operator $S\in 1+{\cal D}_-$ called the wave operator,
and the associated, formally infinite order pseudodifferential operator
$$
	W:= S \E^{\sum_{i=1}^\infty t_i D^i},
$$
as follows:
\begin{equation}
	L=SD S^{-1}=WD W^{-1},
	\label{1.16}
\end{equation}
$$
	\frac{\partial S}{\partial t_n}
	=-(L^n)_-S\,,\quad\mbox{and}\quad
	\frac{\partial W}{\partial t_n}=(L^n)_+W\,.
$$
The wave function
\begin{equation}
	\Psi(t,z):=\Psi(x,t,z):= W\E^{xz}=
	S\E^{\sum_{i=1}^\infty\bar t_iz^i}\,,
	\label{1.18}
\end{equation}
where $\bar t_i=t_i+\delta_{i,1}x$, satisfies
\begin{equation}
	L\Psi=z\Psi\quad\hbox{and}\quad
	\frac{\partial\Psi}{\partial t_n}=(L^n)_+\Psi\,,
	\label{Psi}
\end{equation}
and has the following
representation in terms of
a scalar-valued function associated to $S$ called
the tau function $\tau$:
\begin{eqnarray*}
	\Psi(t,z) &=&\frac{\tau(\bar t-[z^{-1}])}{\tau(\bar t)}
	\E^{\sum_1^\infty\bar t_iz^i}
	\\
	&=&\sum_{n=0}^\infty\frac{p_n(-\tilde\partial)\tau(\bar t)}
	{\tau(\bar t)}z^{-n}\E^{\sum_1^\infty\bar
	t_iz^i}\\
	&=&\sum_{n=0}^\infty\frac{p_n(-\tilde\partial)\tau(\bar t)}
	{\tau(\bar t)}D^{-n}\E^{\sum_1^\infty\bar t_iz^i},
\end{eqnarray*}
implying in view of (\ref{1.18})
\begin{equation}
	S=\frac{\tau(\bar t-[D^{-1}])}{\tau(\bar t)}
	:=\sum_{n=0}^\infty
	\frac{p_n(-\tilde\partial)\tau(\bar t)}{\tau(\bar t)}
	D^{-n}.
\label{WaveOp}
\end{equation}
Moreover, using (\ref{1.18}), we have
$$
	\frac{\partial}{\partial z}\Psi=\frac{\partial}{\partial z}W\E^{xz}=
	W\frac{\partial}{\partial z}\E^{xz} =Wx\E^{xz} =Wx
	W^{-1}\Psi\,,
$$
thus leading to the operator
\begin{eqnarray}
M&:=&	Wx W^{-1}=S\E^{\sum t_kD^k}x \E^{-\sum t_kD^k}S^{-1}
	=S\left(x +\sum_1^\infty kt_kD^{k-1}\right)S^{-1}
	\nonumber\\
&=&	S\left(\sum_1^\infty k\bar t_kD^{k-1}\right)S^{-1} \label{1.27}
\end{eqnarray}
satisfying
$$
	M\Psi=(\partial/\partial z)\Psi\quad\hbox{and}\quad
	[L,M]=W[D,x]W^{-1}=1\,,
$$
and for any formal series $f=f(x,\xi)$,
\begin{equation}
	f(M,L)=Wf(x,D)W^{-1}\,.
	\label{1.29}
\end{equation}

\subsection{Symmetries}

Consider the Lie algebra $w_\infty$ of operators
$$
	w_\infty:=\BC[z,z^{-1}][d/dz]=\Span_\BC\biggl\{
	z^\alpha\biggl(\frac{\partial}{\partial z}\biggr)^\beta
	\biggm|\alpha,\beta\in\BZ, \beta\ge 0\biggr\}\,,
$$
and its completion
$ \overline w_\infty:=\BC((z^{-1}))[\partial/\partial z] $
in the $z^{-1}$-adic topology,
for the customary commutation
relation [ , ]. Acting on $\Psi$,
we have
\begin{equation}
	z^\alpha(\partial/\partial z)^\beta\Psi=M^\beta
	L^\alpha\Psi\,,
\label{1.45}
\end{equation}
motivating the definition of the following
vector fields, called symmetries, on $\Psi$:
$$
	\BY_{z^\alpha(\partial/\partial z)^\beta}\Psi:=
	(M^\beta L^\alpha)_-\Psi\,.
$$
We require that these flows act trivially on parameters $x$, $t$,
and hence on $S^{-1}MS =\sum k\bar t_kD^{k-1}$, for instance.

\begin{lemma}\label{L1.1}
There is an injective homomorphism of Lie algebras
\begin{eqnarray*}
	\overline w_\infty/\BC & \longrightarrow & \left\{
	\begin{tabular}{l}
		Lie algebra of vector fields\\
		on the manifold of wave functions $\Psi$\\
		commuting with the KP flows $\partial/\partial t_n$
	\end{tabular}
	\right\}\\
	z^\alpha\biggl(\frac{\partial}{\partial z}\biggr)^\beta
	& \mapstochar\relbar\joinrel\to
	& \BY_{z^\alpha(\partial/\partial z)^\beta}\Psi =
	(M^\beta L^\alpha)_-\Psi\,,
\end{eqnarray*}
i.e.,
$$
	\left[\BY_{z^\alpha(\partial/\partial z)^\beta},
	\BY_{z^{\alpha'}(\partial/\partial z)^{\beta'}}\right]
	=\BY_{[z^\alpha(\partial/\partial z)^\beta,
	z^{\alpha'}(\partial/\partial z)^{\beta'}]}\,.
$$
\end{lemma}

This definition differs from the one in \cite{asvm} by the sign.
Here this definition is chosen to make it consistent with the
natural action of $\overline w_\infty$ on the Grassmannian discussed in the
next section, rather than its negative.
These vector fields induce
vector fields on $S$ and $L=SDS^{-1}$, as
$$
	\BY_{z^\alpha(\partial/\partial z)^\beta}(S)=
	(M^\beta L^\alpha)_-S
$$
and
$$
	\BY_{z^\alpha(\partial/\partial z)^\beta}(L)=[
	(M^\beta L^\alpha)_-,L]\,.
$$

\begin{proposition}[\cite{asvm}] We have
\begin{equation}
	-\frac{(M^nL^{n+\ell})_-\Psi}{\Psi}=(\E^{-\eta}-1)
	\frac{\frac{1}{n+1}W_\ell^{(n+1)}(\tau)}{\tau}
	\biggr|_{t_1\to t_1+x},
	\quad n,\ell\in\BZ,\ n\ge0,
	\label{asvmformula}
\end{equation}
where the $W^{(n+1)}_\ell$, the generators of the
$W_\infty$-algebra, are the coefficients in the expansion of the
vertex operator
\begin{eqnarray}
	X(t,\lambda,\mu)&:=&
	\exp\left(\sum_{i=1}^\infty(\mu^i-\lambda^i)t_i\right)
	\exp\left(\sum_{i=1}^\infty{\lambda^{-i}-\mu^{-i}\over i}
	\frac{\partial}{\partial t_i}
	\right)\nonumber\\
	&=&\sum_{k=0}^\infty\frac{(\mu-\lambda)^k}{k!}
	\sum_{\ell=-\infty}^\infty\lambda^{-\ell-k}W_\ell^{(k)},
	\quad\mbox{with}\ \
	W_\ell^{(0)}=\delta_{\ell,0}.
	\label{Wgen}
\end{eqnarray}
\end{proposition}

\section{Grassmannian}
\label{sect2}

Let $H:=\BC((z^{-1}))$, $H_+:=\BC[z]$, and $H_-:=z^{-1}\BC[[z^{-1}]]$,
so that $H=H_+\oplus H_-$.
We denote by $\Gr$ the Grassmannian manifold of linear subspaces
${\cal W}$ of $H$ of relative dimension 0 with respect to $H_+$,
i.e., the natural map
$$
	\pi_{\cal W}\colon
	{\cal W}\hookrightarrow H
	\stackrel{\pi}{\longrightarrow} H/H_-\simeq H_+
$$
being Fredholm of index 0.
$\Gr^0:=\{{\cal W}\in\Gr\mid\pi_{\cal W}\hbox{ is isomorphism}\}$
is the big (open) Schubert stratum of $\Gr$.

Given a wave function $\Psi=\Psi(x,t,z)$, let ${\cal W}$ be the
point of $\Gr$ defined by\footnote{
	If $\Psi$ is singular at $(x,t)=0$, we need to replace
	$(\partial/\partial x)^j\Psi(0,0,z)$ in the first line by
	$(\partial/\partial x)^j\bigl(x^n\Psi(x,0,z)\bigr)\bigr|_{x=0}$
	for some $n>0$, and make a similar replacement in the second line
	(see \cite{Sa1} for details).  We chose to write the formulas
	for ${\cal W}\in\Gr^0$ for simplicity.
	\label{excuse}}
\begin{eqnarray*}
	{\cal W}
	&=&\Span_\BC\biggl\{
		\frac{\partial^j}{\partial x^j}\Psi(0,0,z)
		\biggm|j=0,1,2,\ldots
	\biggr\}\\
	&=&\Span_\BC\biggl\{
		{\partial^{j_1+\cdots+j_N}\over
		\partial t_1^{j_1}\ldots\partial t_N^{j_N}}
		\Psi(0,0,z)\biggm|N\ge0,\ j_1,\dots,j_N\ge0
	\biggr\}\,.
\end{eqnarray*}
The first line guarantees ${\cal W}\in\Gr$, and the second line follows
from the first by using the second equation in (\ref{Psi}), i.e., the
KP time evolutions of $\Psi$.
Hence up to the $t$-adic completion we have
$$
	{\cal W}=\Span_\BC\biggl\{
		\biggl(\frac{\partial}{\partial x}\biggr)^j\Psi(0,t,z)
		\biggm| j=0,1,2,\ldots
	\biggr\}\,,
$$
so that, letting $\psi=\E^{-\sum t_iz^i}\Psi$ and
$$
	{\cal W}^t:=\E^{-\sum t_iz^i}{\cal W}
	=\Span_\BC\bigl\{
	(\partial/\partial x)^j\psi(0,t,z)\bigm| j=0,1,2,\ldots\bigr\}\,,
$$
we have $\psi=(\pi_{{\cal W}^t})^{-1}(1)$, i.e., $\psi$ is
the preimage of 1 by the map $\pi_{{\cal W}^t}\colon{\cal W}^t\to H_+$.

The corresponding $\tau$-function $\tau(t)$ is the determinant
of the composite map
\begin{equation}
	{\cal W}\stackrel{g}{\longrightarrow}
	{\cal W}^t\stackrel{\pi_{{\cal W}^t}}{\longrightarrow}
	H/H_-\simeq H_+\,,
	\label{proj}
\end{equation}
where $g$ denotes the multiplication by $\E^{-\sum t_iz^i}$.
Given ${\cal W}$, the determinant is well-defined up to a constant
which is determined by the choice of a basis $\{\psi_k\}_{k=0}^\infty$,
$\psi_k=z^k\bigl(1+O(z^{-1})\bigr)$ for $k\gg0$, of ${\cal W}$.
We take $\{z^k\}_{k=0}^\infty$ as the basis of $H_+$.
More specifically, $\tau(t)$ is defined as the limit as $n\to\infty$ of
the determinant of
\begin{equation}
	{\cal W}_n\hookrightarrow{\cal W}\to H_+\to H_+/z^nH_+\,,
	\label{projn}
\end{equation}
where the middle arrow is the composite map in (\ref{proj}),
${\cal W}_n=\Span_\BC\{\psi_k\}_{k=0}^{n-1}$, and the determinant is
computed with respect to the bases $\{\psi_k\}_{k=0}^{n-1}$ of ${\cal W}_n$
and $\{z^k\}_{k=0}^{n-1}$ of $H_+/z^nH_+$.  The limit exists in the
$t$-adic topology of $\BC[[t]]$, i.e., for any multi-index $\alpha$,
there exists a positive integer $n_\alpha$ such that,
if $n\ge n_\alpha$, then the coefficient of $t^\alpha$ in the determinant of
(\ref{projn}) is independent of $n$, and gives the coefficient
of $t^\alpha$ in $\tau(t)$.  This finiteness property is an immediate
consequence of the fact that, expanding $\tau(t)$ in terms of
Schur functions, the coefficients give the Pl\"ucker coordinates of
${\cal W}$. See \cite{Sa1} for details.

The $\overline w_\infty$-action on $\Psi$ becomes the natural action of
$\overline w_\infty$ on $\Gr$: As an ordinary differential operator in $z$,
each $A\in\overline w_\infty$ acts on $H$, which defines a vector field
on $\Gr$.

\subsection{Stabilizers}

Given ${\cal W}\in\Gr$, we shall call
$$
	S_{\cal W}:=\bigl\{Q:=Q(z,\partial/\partial z)\in\overline w_\infty
	\bigm|Q{\cal W}\subset {\cal W}\bigr\}
$$
the {\em stabilizer\/} of ${\cal W}$.
In this subsection we shall observe basic properties of the stabilizer
which can be obtained without referring to matrix integrals.

\begin{lemma}\label{L2.1} Let ${\cal W}\in\Gr$ and
$A:=\sum_{-\infty<i\ll\infty,0\le j\ll\infty}
c_{ij}z^i(\partial/\partial z)^j\in\overline w_\infty$. If
\begin{equation}
	A{\cal W}\subset {\cal W}\,,
	\label{2.2}
\end{equation}
then
$$
	Q_A:=
	\sum_{\scriptstyle-\infty<i\ll\infty\atop\scriptstyle0\le j\ll\infty}
	c_{ij}M^jL^i\in{\cal D}_+\,.
$$
Conversely, if $Q\in{\cal D}_+$ is of this form, i.e., $Q=Q_A$ for some
$A\in\overline w_\infty$, then this $A$ satisfies (\ref{2.2}).
\end{lemma}

\proof
We have
\begin{equation}
	A\Psi(t,z)=Q_A\Psi(t,z)\label{2.3}
\end{equation}
by definition.
Since $A{\cal W}\subset{\cal W}$, and since the Taylor coefficients
(or Laurent coefficients if ${\cal W}\not\in\Gr^0$)
in $x$ of $\Psi$ generates ${\cal W}$,
$A\Psi$ is a $\BC[[x,t]]$-linear combination of
$\Psi$, $D\Psi$, $D^2\Psi$, \dots, i.e., $A\Psi=Q\Psi$
for some $Q\in{\cal D}_+$.
Hence, since (\ref{2.3}) determines $Q_A$ uniquely, $Q_A$ itself must be in
${\cal D}_+$.
Conversely, suppose $Q_A\in{\cal D}_+$, and let $\Psi(x,0,z)=\sum f_n(z)x^n$
be the Taylor (or Laurent) expansion of $\Psi(x,0,z)$ at $x=0$.
Then each Taylor coefficient in $x$ of $Q_A\Psi$ is a linear combination
of $\{f_n(z)\}$, and hence it belongs to
${\cal W}$, so that by (\ref{2.3}) $Af_n\in{\cal W}$ for every $n$
(the action of $A$ on $f_n$ is well-defined since $A$ is a differential
operator in $z$).  Since $\{f_n\}$ is a basis of ${\cal W}$, we have
$A{\cal W}\subset{\cal W}$.
\endproof

\begin{corollary}
Let $p\ne0$ be an integer, and let $Q\in{\cal D}_+$ such that
$\ad(L^p)^NQ=0$ for $N\gg0$.  Then $Q=Q_A$ for some $A\in\overline w_\infty$
such that $A{\cal W}\subset{\cal W}$ holds.
In particular, a solution to the string equation (\ref{0.1}) always
comes from a pair of $A\in\overline w_\infty$ and ${\cal W}\in\Gr$, such
that $A{\cal W}\subset{\cal W}$ (and $z^p{\cal W}\subset{\cal W}$
due to the extra assumption $P=L^p\in{\cal D}_+$).
\end{corollary}

\proof
Writing $Q=\sum_{ij}c_{ij}M^jL^i$, let
$A=\sum_{ij}c_{ij}z^i(\partial/\partial z)^j$.
Since $\ad(L^p)^NQ=0$ we have $\ad(z^p)^NA=0$, which implies that $A$ is a
differential operator in $z$.  Hence the ``converse" part of
Lemma~\ref{L2.1} applies.
\endproof

\begin{lemma}\label{three}
Let $A$,~$B\in\overline w_\infty$,
$\psi_0=1+O(z^{-1})\in 1+H_-$ and ${\cal W}\in\Gr$.
Suppose $A$ acts on the monomials $z^k$, $k\in\BZ$, as
$$
	Az^k=z^{k+1}\bigl(c_k+O(z^{-1})\bigr)\,,
$$
and $c_k\ne0$ if $k\ge0$.  Then the following conditions are equivalent:
\begin{itemize}
\item[(i)]	$\psi_0\in{\cal W}$, $A{\cal W}\subset{\cal W}$ and
		$B{\cal W}\subset{\cal W}$;
\item[(ii)]	${\cal W}=\Span_\BC\{\psi_0,A\psi_0,A^2\psi_0,\ldots\}$, and
		$\psi_0$
		satisfies the differential equations
		\begin{equation}
			BA^n\psi_0=F_n(A)\psi_0\,,\quad n=0,1,\ldots
		\label{2.5}
		\end{equation}
		for some $F_n(s)\in\BC[s]$.
\end{itemize}
In particular, under these conditions ${\cal W}$ belongs to the big stratum
$\Gr^0$ of $\Gr$.
If, moreover, $A$ and $B$ satisfy a commutation relation of the form
\begin{equation}
	[A,B]=a(A)B+b(A)
	\label{2.6}
\end{equation}
for some $a(s)$,~$b(s)\in\BC[s]$, then in (\ref{2.5}) it suffices to assume
only the $n=0$ case, i.e.,
\begin{equation}
	B\psi_0=F(A)\psi_0
	\label{2.7}
\end{equation}
for some $F(s)\in\BC[s]$.
\end{lemma}

\proof
Since $\psi_0\in{\cal W}$, $A{\cal W}\subset{\cal W}$ implies
$
	{\cal W}'
	:=\Span_\BC\{\psi_0,A\psi_0,A^2\psi_0,\ldots\} \subset {\cal W}
$.
Since $\psi_0=1+O(z^{-1})$ and
$A$ raises the order of a function in $z$ by 1, the map ${\cal W}'\to H_+$
is bijective, and ${\cal W}'\in\Gr^0$.  In particular, both ${\cal W}$ and
${\cal W}'$ are of relative dimension 0, so that ${\cal W}={\cal W}'$.
Conversely, ${\cal W}={\cal W}'$ clearly implies $\psi_0\in{\cal W}$ and
$A{\cal W}\subset{\cal W}$.
Assume these equivalent conditions.  Then $B{\cal W}\subset{\cal W}$ if and
only if $B{\cal W}'\subset{\cal W}'$ if and only if the differential equations
of the form (\ref{2.5}) are satisfied.  Finally, when $A$ and $B$ satisfy a
commutation relation of the form (\ref{2.6}), the $n^{\hbox{\scriptsize th}}$
equation in (\ref{2.5}) implies the $(n+1)^{\hbox{\scriptsize st}}$ one,
so that (\ref{2.7}) suffices.
\endproof

The following propositions take a closer look at the $[P,Q]=1$ case and
$[P,Q]=P$ case, to show that essentially those elements in
$\overline w_\infty$ which give rise to $P$ and $Q$ in the sense of
Lemma~\ref{L2.1}, and their polynomials, are the only elements of
the stabilizer.

\begin{proposition}\label{stable1}
Let $p\in\BZ$, $p>0$.
Let $A\in\overline w_\infty$ be such that $[A,z^p]=1$.
If ${\cal W}\in\Gr$ satisfies $z^p{\cal W}\subset{\cal W}$ and
$A{\cal W}\subset{\cal W}$,
then the stabilizer of ${\cal W}$ is generated by $z^p$ and $A$, i.e.,
$S_{\cal W}=\BC[A,z^p]$.
\end{proposition}

\proof
Since $[A,z^p]=1$, $A$ is a first order
differential operator in $z$, so that any $C\in S_{\cal W}$ can be written
as $C=\sum_{-\infty<i\ll\infty,0\le j\le N} a_{ij}z^i A^j$ for some $N\ge0$.
It suffices to prove that $a_{ij}=0$ if $i<0$ or if $i\not\equiv0\bmod p$.
Suppose $A$ raises the order of a function in $z$ by $k$:
$\ord_zAz^\ell=\ell+k$.  Let
$I$ be the set of pairs $(i,j)$ such that $i<0$ or $i\not\equiv0\bmod p$,
$a_{ij}\ne0$, and $i+kj$ is maximum among all such $a_{ij}$'s.  We have
$|I|<\infty$, and we only need to prove $|I|=0$.  Suppose this is not true.
Let $C_0:=\sum_{(i,j)\in I} a_{ij}z^i A^j$.
Noting
$$
	[A,z^iA^j]=[A,(z^p)^{i/p}]A^j=(i/p)z^{i-p}A^j,
$$
so that $\ad(A)^n(z^iA^j)=0$ for $n\gg0$ if and only if
$i\ge0$ and $i\equiv0\bmod p$, we see that for $n\gg0$
the leading terms of $\ad(A)^nC$ are $\ad(A)^nC_0$, which
lowers the order of a function in $z$, and does not annihilate
the function for a general $n$.  This cannot happen
since $\ad(A)^nC{\cal W}\subset{\cal W}$, and since in ${\cal W}$
the order of functions in $z$ are bounded from below.
\endproof

\begin{proposition}\label{stable2}
Let $p\in\BZ$, $p>0$.
Let $A=z\partial/\partial z-a(z)$, where $a(z)\in z+\BC[[z^{-1}]]$, and
$\psi_0=1+O(z^{-1})\in 1+H_-$.
Let ${\cal W}\in\Gr$ be the point of the Grassmannian determined by
the conditions $\psi_0\in{\cal W}$ and $A{\cal W}\subset{\cal W}$.
Suppose ${\cal W}$ also satisfies $z^p{\cal W}\subset{\cal W}$.
Let $F(s)=c\prod_{i=1}^p(s-c_i)\in\BC[s]$, where
$c_i\in\BC$, $c\in\BC^*$, be the polynomial of degree $p$ as in
(\ref{2.7}) with $B=z^p$, i.e., $\psi_0$ satisfies the equation
\begin{equation}
	F(A)\psi_0=z^p\psi_0\,.
\label{S1}
\end{equation}
Then if $F$ satisfies the following genericity condition:
\begin{itemize}
\item[(G)]
	For any $n\not\equiv0\bmod p$, we have
	$(F)+n:=\sum(c_i+n)\not\equiv(F)\bmod p$, i.e.,
	$\pi_p\bigl((F)+n\bigr)\ne\pi_p(F)$,
	where $(F)=\sum_{i=1}^p(c_i)$ is the divisor of $F$,
	and $\pi_p\colon\BC\to\BC/p\BZ$
	is the natural projection,
\end{itemize}
then the stabilizer of ${\cal W}$
is generated by $A$, $z^p$ and $\xi:=z^{-p}F(A)$, i.e.,
\begin{equation}
	S_{\cal W}=\BC[A,z^p,\xi]\,.
\label{S2}
\end{equation}
\end{proposition}

\begin{remark}
Condition (G) is equivalent to
\begin{itemize}
\item[(G$'$)]
	There does {\em not\/} exist $n\mid p$, $0<n<p$, and $H(s)\in\BC[s]$
	of degree $n$ such that $F(s)=\prod_{i=0}^{p/n}H(s-in)$;
\end{itemize}
and if it is not satisfied, i.e., if $F(s)=\prod_{i=0}^{p/n}H(s-in)$
for some $n\mid p$ and $H$, then taking such $(n,H)$ of the smallest $n$,
we observe from our proof below that
$\BC[A,z^p,\xi]\subset S_{\cal W}\subset\BC[A,z^n,\xi']$,
where $\xi'=z^{-n}H(A)$.
\end{remark}
\begin{remark}
The right-hand side of (\ref{S2}) equals
$\sum_{i,j,k\ge0}\BC a^ib^jc^k$,
where $(a,b,c)$ is any permutation of $(A,z^p,\xi)$;
the order does not matter because
\begin{equation}
	[A,z^p]=pz^p\,,\quad [A,\xi]=-pF(A)\quad\hbox{and}\quad
	[z^p,\xi]=F(A)-F(A-p)\,.
\label{S3}
\end{equation}
\end{remark}
\begin{remark}\label{R3}
Condition (G) is satisfied by the $F$ in
Theorem~\ref{T2}: Since
$$
	F(s)=\left(\prod_{i=0}^{p-2}(s-i)\right)\bigl(s-(p-1+cp)\bigr)
$$
and $-1<c<0$, there is no period less than $p$ in the divisor of $F$
modulo $p$.
\end{remark}

\medbreak\noindent{\em Proof of Prop.~\ref{stable2}}.
Using the commutation relations
(\ref{S3}), the definition of ${\cal W}$, and Eq.~(\ref{S1}),
we observe easily that $S_{\cal W}\supset\BC[A,z^p,\xi]$.
We prove the converse inclusion in two steps.
Only Step~2 needs Condition (G).
\smallskip\par\noindent
{\it Step 1}. We observe that $S_{\cal W}$ is spanned by the $z$-homogeneous
elements in $S_{\cal W}$, i.e., the elements of $S_{\cal W}$ of the form
$z^nf(A)$, where $n\in\BZ$ and $f(s)\in\BC[s]$.

Indeed, let $S'\subset S_{\cal W}$ be the subspace of $S_{\cal W}$
spanned by the $z$-homogeneous elements, and suppose that
$S'':=S_{\cal W}\setminus S'\ne\emptyset$.
Let $N$ be a nonnegative integer such that
$$
	S''{}^{(N)}:=\bigl\{
	C\in S''\bigm|\ord_{\partial/\partial z}C\le N\bigr\}
$$
is nonempty.  Let $C\in S''{}^{(N)}$ be such that, writing
\begin{equation}
	C=\sum z^nf_n(A)\,,
\label{S4}
\end{equation}
$n_0(C):=\max\{n\mid f_n\not\equiv0\}$ is the smallest in $S''^{(N)}$.
Such a $C$ exists because
\begin{claim}
$\bigl\{n_0(C)\bigm|C\in S''^{(N)}\bigr\}$ is bounded below.
\end{claim}
\proof
Indeed it is bounded from below by $-2N+1$:
since $C\in S''^{(N)}$ is an ordinary differential operator
of $\hbox{order}\le N$, and since $\psi_0$, $A\psi_0,\dots,A^{N-1}\psi_0$
are linearly independent, we have $C A^i\psi_0\not\equiv0$ for some
$i$, $0\le i<N$. Since $A^i\psi_0=(-1)^iz^i\bigl(1+O(z^{-1})\bigr)$,
since $C{\cal W}\subset{\cal W}$, and since ${\cal W}$ is a span of
$A^j\psi_0$ for $j\ge0$, we observe that $C$ does not decrease the order
of $A^i\psi_0$ in $z$ by more than $N-1$. This implies, using
the notation of (\ref{S4}), that $n+\deg f_n\ge-(N-1)$ for some
$n$.  Hence $n_0(C)\ge n\ge -\deg f_n-(N-1)\ge-(2N-1)$.
\endproof

Now let
\begin{equation}
	C':=[A,C]-n_0(C)C=\sum\bigl(n-n_0(C)\bigr)z^nf_n(A)\,.
\label{S5}
\end{equation}
Clearly $C'\in S_{\cal W}$. We have
$\ord_{\partial/\partial z}C'\le\ord_{\partial/\partial z}C\le N$,
and $n_0(C')\le n_0(C)-1$. Hence by the minimality of $n_0(C)$,
we must have $C'\not\in S''{}^{(N)}$, so that $C'\in S'$.
Thus each term $\bigl(n-n_0(C)\bigr)z^nf_n(A)$ in (\ref{S5}) belongs
to $S'$, and
only finitely many $f_n$ are non-zero. As a finite linear combination
of such, we have $C'':=C-z^{n_0(C)}f_{n_0(C)}(A)\in S'$,
so that $z^{n_0(C)}f_{n_0(C)}(A)=C-C''$ must also belong to $S_{\cal W}$,
and hence to $S'$, since it is $z$-homogeneous.
This implies $C=C''+(C-C'')\in S'$, which is a contradiction.
\smallskip\par\noindent
{\it Step 2}. Let $f(s)\not\equiv0$ be any constant coefficient
polynomial, and let $n$ be an integer. We prove that
$$
	z^nf(A)\in S_{\cal W}\quad\hbox{implies}\quad p\mid n\,,
$$
and that, when $n<0$, $z^nf(A)\in S_{\cal W}$ must have the form $\xi^kh(A)$
for $k:=-n/p>0$ and some $h(s)\in\BC[s]$.

Suppose $z^nf(A)\in S_{\cal W}$. We assume $n\ne0$ without loss of
generality.  Since $z^nf(A)\psi_0\in{\cal W}$, by Lemma~\ref{three}
there exists another
polynomial $g(s)\in\BC[s]$, such that
\begin{equation}
	z^nf(A)\psi_0=g(A)\psi_0\,.
\label{S6}
\end{equation}

First assume $n>0$.
Let $\ell>0$ be the least common multiple of $p$ and $n$.
Noting
$$
	z^{2p}\psi_0=z^pF(A)\psi_0=F(A-p)z^p\psi_0=F(A-p)F(A)\psi_0
$$
etc., we have
\begin{equation}
	\left(\prod_{i=0}^{\ell/p-1}F(A-ip)\right)\psi_0=z^\ell\psi_0
\label{S7}
\end{equation}
from (\ref{S1}), and
\begin{equation}
	\left(\prod_{j=0}^{\ell/n-1}G(A-jn)\right)\psi_0=z^\ell\psi_0
\label{S8}
\end{equation}
from (\ref{S6}), where $G(s)=g(s)/f(s-n)$ is a rational function in $s$,
and $G(A-jn)$ in (\ref{S8}) is understood as an element of the field of
fractions of $\BC[A]$; this makes sense because, since
$\{A^n\psi_0\}_{n=0,1,\dots}$ is linearly independent, the representation
$$
	\BC[s]\ni f(s)\mapsto f(A)\psi_0\in{\cal W}
$$
is faithful.

Comparing the left-hand sides of (\ref{S7}) and (\ref{S8}), we thus have
the equality
\begin{equation}
	\prod_{i=0}^{\ell/p-1}F(s-ip)=\prod_{j=0}^{\ell/n-1}G(s-jn)
\label{S9}
\end{equation}
of rational functions in $s$.
Since the left-hand side of it is a polynomial of $s$, so is the right-hand
side. Let $D$ be the divisor of this polynomial, and let $\pi_\ell$
be the natural map $\BC\to\BC/\ell\BZ$. From the left-(resp.\ right-)hand
side of (\ref{S9}) the image $\pi_{\ell}(D)$ of divisor $D$
on the cylinder
$\BC/\ell\BZ$ is invariant under the translation by $p$ (resp.\ $n$).
But the genericity condition (G) implies that if $\pi_{\ell}(D)$ is
invariant under the translation by $k\in\BZ$, then $p\mid k$.
Hence $p\mid n$.

Note here that, since $\ell$ is the least common multiple of $p$ and $n$,
this implies $\ell=n$, so that the right-hand side of
(\ref{S9}) is $G(s)$ itself.  Hence
$$
	g(s)/f(s-n)=G(s)=\prod_{i=0}^{n/p-1}F(s-ip)\,.
$$
In particular, $g(s)/f(s-n)$ is a polynomial.

In the case where $n<0$, after rewriting (\ref{S6}) as
$$
	z^{-n}g(A)\psi_0=f(A)\psi_0\,,
$$
we switch the roles of $f$ and $g$, and $n$ and $-n$, to
proceed exactly the same way to prove $p\mid n$ and
$$
	f(s)/g(s+n)=\prod_{i=0}^{-n/p-1}F(s-ip)\,.
$$
Thus we have
\begin{eqnarray*}
	z^nf(A)
	&=&z^n\Biggl(\prod_{i=0}^{-n/p-1}F(A-ip)\Biggr)g(A+n)\\
	&=&\bigl(z^{-p}F(A)\bigr)^{-n/p}g(A+n)\\
	&=&\xi^kg(A+n)=:\xi^kh(A)\,,
\end{eqnarray*}
proving the last assertion of Step 2, and hence completing the proof of
Prop.~\ref{stable2}.

\subsection{Symmetric functions and matrix integrals}

In this subsection, we prove a number of lemmas regarding symmetric functions.

\begin{lemma}\label{L2.3}
Let $s$ and $N$ be positive integers.
Let $F\bigl(x^{(1)},\ldots,x^{(s)}\bigr)$ be a function which is symmetric
in each $x^{(r)}:=\bigl(x_1^{(r)},\dots,x_N^{(r)}\bigr)\in\BC^N$,
$r=1,\ldots,s$; let $f_1,\ldots,f_s$ be functions of two variables,
and let $B\bigl(x^{(s)}\bigr)$ be a skew-symmetric function of $x^{(s)}$.
If $C_1,\dots,C_s$ denote $s$ fixed contours in $\BC$, then the integral
\begin{eqnarray*}
\Phi\bigl(x^{(0)}\bigr)
&:=&	\int\cdots\int_{(C_1)^N\times\cdots\times(C_s)^N}
	\prod_{r=1}^s\prod_{i=1}^N dx^{(r)}_i\cdot \\
&&	\cdot F\bigl(x^{(1)},\ldots,x^{(s)}\bigr)B\bigl(x^{(s)}\bigr)
	\prod_{r=1}^s
	\det\left(f_r\bigl(x_i^{(r-1)},x_j^{(r)}\bigr)\right)_{1\le i,j\le N}
	\,,
\end{eqnarray*}
where $x^{(0)}\in\BC^N$ comes in as the first argument of $f_1$,
is skew-symmetric in $x^{(0)}$, and
\begin{eqnarray*}
\Phi\bigl(x^{(0)}\bigr)
&=&	(N!)^s\int\cdots\int_{(C_1)^N\times\cdots\times(C_s)^N}
	\prod_{r,i}dx^{(r)}_i\cdot \\
&&	\cdot F\bigl(x^{(1)},\ldots,x^{(s)}\bigr)B\bigl(x^{(s)}\bigr)
	\prod_{r=1}^s
	\prod_{i=1}^N f_r\bigl(x_i^{(r-1)},x_i^{(r)}\bigr)\,.
\end{eqnarray*}
\end{lemma}

\proof
For any (good) functions $A=A\bigl(x^{(1)},\ldots,x^{(s)}\bigr)$
and $h=h\bigl(x^{(1)},\ldots,x^{(s)}\bigr)$, let
$$
	\langle Ah\rangle:=\int\cdots\int_{(C_1)^N\times\cdots\times(C_s)^N}
	\prod_{r,i}dx^{(r)}_i\cdot
	A\bigl(x^{(1)},\ldots,x^{(s)}\bigr)h\bigl(x^{(1)},\ldots,x^{(s)}\bigr)
	\,.
$$
For any $\sigma_r\in{\frak S}_N$, let
$x^{(r)}_{\sigma_r}:=
\bigl(x^{(r)}_{\sigma_r1},\ldots,x^{(r)}_{\sigma_rN}\bigr)$, and
$h^{(\sigma_1,\ldots,\sigma_s)}\bigl(x^{(1)},\ldots,x^{(s)}\bigr)
:=h\bigl(x^{(1)}_{\sigma_1},\ldots,x^{(s)}_{\sigma_s}\bigr)$.
Clearly $\langle Ah\rangle = \bigl<
A^{(\sigma_1,\ldots,\sigma_s)}h^{(\sigma_1,\ldots,\sigma_s)}\bigr>$.
If, moreover, $A$ is symmetric in each of $x^{(1)},\ldots,x^{(s-1)}$,
and skew-symmetric in $x^{(s)}$, i.e.,
$A^{(\sigma_1,\ldots,\sigma_s)}=(-1)^{\varepsilon(\sigma_s)}A$,
then we have
$$
	\langle Ah\rangle = \bigl<
	A^{(\sigma_1,\ldots,\sigma_s)}h^{(\sigma_1,\ldots,\sigma_s)}
	\bigr> = (-1)^{\varepsilon(\sigma_s)}
	\bigl< Ah^{(\sigma_1,\ldots,\sigma_s)}\bigr>
	\quad\forall \sigma_r\in{\frak S}_N\,.
$$
Applying this to
$h\bigl(x^{(1)},\ldots,x^{(s)}\bigr):=
\prod_r\prod_i f_r\bigl(x^{(r-1)}_i,x^{(r)}_i\bigr)$, and summing it up over
$(\sigma_1,\ldots,\sigma_s)\in({\frak S}_N)^s$, we obtain
\begin{eqnarray*}
\lefteqn{(N!)^s\biggl<
		A\prod_r\prod_i f_r\bigl(x^{(r-1)}_i, x^{(r)}_i\bigr)
	\biggr>}\\
&=&	\biggl<A\sum_{\sigma_1,\ldots,\sigma_s}(-1)^{\varepsilon(\sigma_s)}
	\prod_r\prod_i
	f_r\bigl(x^{(r-1)}_{\sigma_{r-1}i},x^{(r)}_{\sigma_ri}\bigr)
	\biggr>\,,\quad\hbox{with $\sigma_0 = \rm id$,}\\
&=&	\biggl<A\sum_{\sigma_1,\ldots,\sigma_s} \prod_r
	(-1)^{\varepsilon(\sigma_r)-\varepsilon(\sigma_{r-1})}
	\prod_i f_r\bigl(x^{(r-1)}_{\sigma_{r-1}i},x^{(r)}_{\sigma_ri}\bigr)
	\biggr>\\
&=&	\biggl<A \prod_r\sum_{\sigma\in{\frak S}_N}(-1)^{\varepsilon(\sigma)}
	\prod_i f_r\bigl(x^{(r-1)}_i,x^{(r)}_{\sigma i}\bigr)
	\biggr>\\
&=&	\biggl<A\prod_r
	\det\left(f_r\bigl(x^{(r-1)}_i,x^{(r)}_j\bigr)\right)_{i,j}
	\biggr>\,.
\end{eqnarray*}
Setting here $A=F\bigl(x^{(1)},\ldots,x^{(s)}\bigr)B\bigl(x^{(s)}\bigr)$
proves the identity in Lemma~\ref{L2.3}.
Finally, $\Phi\bigl(x^{(0)}\bigr)$ is skew-symmetric in $x^{(0)}$ since
$\det\left(f_1\bigl(x^{(0)}_i,x^{(1)}_j\bigr)\right)$ is.
\endproof

\begin{lemma}
\label{L2.4}
(See \cite[Lemma~4.2]{kon}, \cite[Eq.~(2.21)]{GKM}, \cite[Theorem~8.18]{mul}.)
Let
$$
	{\cal W}=\Span_\BC\bigl\{
	\psi_0(z),\psi_1(z),\psi_2(z),\ldots\bigr\}\in\Gr
$$
with functions
$$
	\psi_k(z)=\sum_{-\infty<j\le k}a_{j,k}z^j,\quad\quad k=0,1,2,\dots,
$$
such that $a_{kk}=1$ for $k\gg0$, i.e., $\ord_z\psi_k(z)\le k$,
and $\psi_k(z)=z^k\bigl(1+O(z^{-1})\bigr)$ for $k\gg0$.
Let $N>0$ be any integer such that this condition holds for $k\ge N$.
Let $z_1,\dots,z_N$ be formal scalar variables near $\infty$. Then
the $\tau$-function $\tau(t)$ at
\begin{equation}
	t_n:=-\frac{1}{n}\sum^N_{i=1}z_i^{-n},\quad n=1,2,\dots,
	\label{miwa}
\end{equation}
is given by
\begin{equation}
	\tau(t)=\frac{\det\bigl(\psi_{j-1}(z_i)\bigr)_{1\le i,j\le
	N}}{\det(z_i^{j-1})_{1\le i,j\le N}}\,.
	\label{formula}
\end{equation}
\end{lemma}

\proof
Our proof is based on Kontsevich's idea in \cite{kon};
see \cite[Sect.~2.3]{GKM} for a proof using free fermions.
To keep the notation simple, let us denote by $(1-z)^{-1}$ and $(-z+1)^{-1}$
the geometric series $\sum_0^\infty z^n$ and $-\sum_{-\infty}^{-1}z^n$,
respectively.
Let $\delta(z):=(1-z)^{-1}-(-z+1)^{-1}=\sum_{-\infty}^\infty z^n$,
which plays the role of delta function, in the sense that
\begin{equation}
	\delta(z/y)f(z)=\delta(z/y)f(y)\,,
	\label{delta}
\end{equation}
as is obvious by taking $f(z)=z^m$ (see \cite{djkm}).
Let $\sigma:=\prod_{i=1}^N(-z_i)=(-1)^Nz_1\ldots z_N$.
Let $\sigma_i:=1\big/\prod_{j(\ne i)}(1-z_i/z_j)$, $i=1,\dots,N$,
understood as rational functions of $z_j$'s, so that we have
the following identity of formal power series in $z$:
$$
	\prod_{i=1}^N\left(1-{z\over z_i}\right)^{-1}
	=\sum_{i=1}^N\sigma_i\left(1-{z\over z_i}\right)^{-1}.
$$
 From (\ref{miwa}) we have
\begin{eqnarray*}
g:=\exp\left(-\sum_{n=1}^\infty t_nz^n\right)
	&=&\prod_{i=1}^N\left(1-{z\over z_i}\right)^{-1}
	=\sum_{i=1}^N\sigma_i\left(1-{z\over z_i}\right)^{-1} \\
	&=&\sum_{i=1}^N\sigma_i\delta(z/z_i)
	+\sum_{i=1}^N\sigma_i\left(-{z\over z_i}+1\right)^{-1} \\
	&=&\sum_{i=1}^N\sigma_i\delta(z/z_i)
	+\prod_{i=1}^N\left(-{z\over z_i}+1\right)^{-1}\,,
\end{eqnarray*}
so that by using (\ref{delta}), we have
\begin{eqnarray*}
	g\psi_j(z)
	&=&\sum_{i=1}^N\sigma_i\delta(z/z_i)
	\psi_j(z)
	+\left(\prod_{i=1}^N\left(-{z\over z_i}+1\right)^{-1}\right)
	\psi_j(z)
	\\
	&=&\sum_{i=1}^N\sigma_i\delta(z/z_i)
	\psi_j(z_i)
	+ z^{-N}\bigl(\sigma+O(z^{-1})\bigr)
	\psi_j(z)\,.
\end{eqnarray*}
Denoting by $B$ the matrix of the composite map in (\ref{proj}) with respect
to the bases $\{\psi_j\}_{j=0}^\infty$ and $\{z^k\}_{k=0}^\infty$,
we have thus $B=B^0+B^1$, where
$$
	B^0=\left(\begin{array}{ccc}
		1 & \cdots & 1\\
		z_1^{-1} & \cdots & z_N^{-1}\\
		z_1^{-2} & \cdots & z_N^{-2}\\
		\vdots   & \cdots & \vdots
	\end{array}\right)
	S_N
	\left(\begin{array}{ccc}
		\psi_0(z_1) & \psi_1(z_1) & \cdots\\
		\vdots      & \vdots      & \vdots\\
		\psi_0(z_N) & \psi_1(z_N) & \cdots
	\end{array}\right),
$$
$$
	B^1=\left(\begin{array}{ccc|cccc}
		\cdots&0&0 &\smash{\overbrace{0\ \cdots\ 0}^N}&\sigma&&*\\
		\cdots&0&0 &0\ \cdots\ 0&0&\sigma&\\
		\cdots&\vdots&\vdots&\vdots\ \cdots\ \vdots&\vdots&
		\hss\ddots\hss&\hss\ddots\\
		\cdots&\vdots&\vdots&\vdots\ \cdots\ \vdots&\vdots&
		\hss\cdots\hss&\hss\ddots\\
	\end{array}\right)
	\left(\begin{array}{ccc}
		\vdots&\vdots&\cdots\\
		a_{-2,0}&a_{-2,1}&\cdots\\
		a_{-1,0}&a_{-1,1}&\cdots\\
		\hline
		a_{00}&a_{01}&\cdots\\
		0&a_{11}&\cdots\\
		0&0&\ddots\\
		\vdots&\vdots&\ddots
	\end{array}\right),
$$
$S_N$ is the diagonal matrix $\diag(\sigma_1,\dots,\sigma_N)$,
and $a_{kj}$, $-\infty<k<\infty$, $0\le j<\infty$, are the Laurent
coefficients of $\psi_j=\sum_k a_{kj}z^k$.

Let us apply some column operations on $B$.  Adding an appropriate linear
combination of first $N$ columns to the $(N+i)^{\hbox{\scriptsize th}}$
column ($i>0$), we can eliminate the column
${^t}\bigl(\psi_{N+i}(z_1),\dots,\psi_{N+i}(z_N)\bigr)$, $i>0$, from $B^0$.
Since $N$ is large enough so that $a_{jj}=1$ for $j\ge N$,
$B^1$ has the form
$$
	\left(O_{\infty\times N}\ \begin{array}{|ccc}
		\sigma & & * \\
		&\sigma&\\
		&&\ddots\\
		\hbox{\Large 0}&&
	\end{array}\right),
$$
so that the ``$*$" part can be eliminated by further column operations on
columns $N+1$, $N+2$,~\dots, which do not alter the $B^0$-part.
Here $O_{m\times n}$ is the $m\times n$ zero matrix.
The matrix $B$ can thus be reduced to $B'=B'^0+B'^1$, where
$$
	B'^0=\left(\begin{array}{ccc}
		1 & \cdots & 1\\
		z_1^{-1} & \cdots & z_N^{-1}\\
		z_1^{-2} & \cdots & z_N^{-2}\\
		\vdots   & \cdots & \vdots
	\end{array}\right)
	S_N
	\left(\begin{array}{ccc|}
		\psi_0(z_1) & \cdots &\psi_N(z_1)\\
		\vdots      & \vdots&\vdots\\
		\psi_0(z_N) & \cdots &\psi_N(z_N)
	\end{array}
	\ O_{N\times\infty}\right),
$$
$$
	B'^1=\Bigl(O_{\infty\times N}\Bigm|\sigma I_\infty\Bigr)\,.
$$
Let $n$, $n\ge N$, be an integer.
Note that the column operations needed to bring $B$ into $B'$ only adds
linear combinations of lower numbered columns to higher ones.  Hence,
denoting by $B_n$, $B'_n$, $B'^0_n$ and $B'^1_n$ the matrices of the first
$n$ rows and columns in $B$, $B'$, $B'^0$ and $B'^1$, respectively, we have
$\det B_n =\det B'_n=\det(B'^0_n + B'^1_n)$, with
$$
	B'^0_n=\left(\begin{array}{ccc}
		1 & \cdots & 1\\
		z_1^{-1} & \cdots & z_N^{-1}\\
		\vdots   & \cdots & \vdots\\
		z_1^{-n+1} & \cdots & z_N^{-n+1}
	\end{array}\right)
	S_N
	\left(\begin{array}{ccc|}
		\psi_0(z_1) & \cdots &\psi_N(z_1)\\
		\vdots      & \vdots&\vdots     \\
		\psi_0(z_N) & \cdots &\psi_N(z_N)
	\end{array}
	\ O_{N\times(n-N)}\right)\,,
$$
and
$$
	B'^1_n=\left(\begin{array}{c|c}
		O_{(n-N)\times N} & \sigma I_{n-N}\\
		\hline
		O_{N\times N} & O_{N\times(n-N)}
	\end{array}\right)\,.
$$
Since the last $n-N$ columns of $B'^0_n$ are 0, we have
$$
	B'_n=\left(\begin{array}{c|c}
		* & \sigma I_{n-N}\\
		\hline
		Z & O_{N\times(n-N)}\end{array}
	\right),
$$
where $Z$
consists of the last $N$ rows and the first $N$ columns of $B'^0_n$:
$$
	Z=\left(\begin{array}{ccc}
		z_1^{-n+N} & \cdots & z_N^{-n+N}\\
		\vdots   & \cdots & \vdots\\
		z_1^{-n+1} & \cdots & z_N^{-n+1}
	\end{array}\right)
	S_N
	\left(\begin{array}{ccc}
		\psi_0(z_1) & \cdots &\psi_N(z_1)\\
		\vdots      & \vdots&\vdots     \\
		\psi_0(z_N) & \cdots &\psi_N(z_N)
	\end{array}
	\right)\,.
$$
Hence we have, using $\sigma=(-1)^Nz_1\ldots z_N$,
\begin{eqnarray*}
\det B_n=\det B'_n
	&=&(-1)^{N(n-N)}\det Z\det(\sigma I_{n-N})\\
	&=&(z_1\ldots z_N)^{n-N}\det Z\\
	&=&(z_1\ldots z_N)^{1-N}\det Z'\,,
\end{eqnarray*}
where
$$
	Z'=\left(\begin{array}{ccc}
		z_1^{N-1} & \cdots & z_N^{N-1}\\
		\vdots   & \cdots & \vdots\\
		z_1^1 & \cdots & z_N^1\\
		1 & \cdots & 1
	\end{array}\right)
	S_N
	\left(\begin{array}{ccc}
		\psi_0(z_1) & \cdots &\psi_N(z_1)\\
		\vdots      & \vdots&\vdots     \\
		\psi_0(z_N) & \cdots &\psi_N(z_N)
	\end{array}
	\right)\,.
$$
Noticing
$$
	\det(z_j^{N-i})_{1\le i,j\le N}=
	(-1)^{N(N-1)/2}\det(z_j^{i-1})_{1\le i,j\le N},
$$
and
$$
	\det S_N=\prod_1^N\sigma_i
	={\left(\prod_{j=1}^Nz_j\right)^{N-1}\over
	\prod_{i,j\ne i}(z_j-z_i)}
	={(z_1\ldots z_N)^{N-1}\over
	(-1)^{N(N-1)/2}\det(z_j^{i-1})^2_{1\le i,j\le N}}\,,
$$
we observe that $\det B_n$ coincides with the right-hand side of
(\ref{formula}). Since $n\ge N$ is arbitrary, this completes the proof
of Lemma~\ref{L2.4}.
\endproof

\begin{lemma}\label{L2.5}
Let $Z:=\diag(z_1,\dots,z_N)$.
Let $\lambda:=\bigl((p-1)(N-1),(p-1)(N-2),\ldots,p-1\bigr)$.
For a polynomial $f(y,z)$, let us denote by $\bigl(f(y,z)\bigr)_2$
the terms in $f(y,z)$ which are quadratic in $y$.
Then we have\footnote{
	$F_\lambda$ is the Schur function for the
	partition $\lambda$.}
\begin{eqnarray*}
	\frac{\Delta(z^p)}{\Delta(z)}&=&F_\lambda\left(
		-\tr Z,-\frac{1}{2}\tr Z^2, -\frac{1}{3}\tr Z^3,\ldots
	\right) \\
	&=&c\prod z_i^{-\frac{p-1}{2}}\left(\int_{{\cal H}_N} dY\exp\,
	\tr\left(-\frac{(Y+Z)^{p+1}}{p+1}\right)_2\right)^{-1},
\end{eqnarray*}
where $c$ is a non-zero constant which depends only on $N$ and $p$.
\end{lemma}

\proof
The Schur function associated with the partition $\lambda$ is given by
(see \cite{Macdonald})
$$
	F_\lambda\left(
		-\sum_1^N y_i, -\frac{1}{2}\sum_1^N y_i^2,
		-\frac{1}{3}\sum_1^N y_i^3,\ldots
	\right):=
	\frac{\Delta_{\lambda+\delta}(y)}{\Delta_\delta(y)}\,,
$$
where $\delta=(N-1>N-2>\cdots>1>0)$ and
$\Delta_\mu(y)=\det(y_i^{\mu_j})_{1\le i,j\le N}$.
Therefore we have,
with $\lambda+\delta= \bigl(p(N-1)> p(N-2)>\cdots>p>0\bigr)$,
$$
	\frac{\Delta(z^p)}{\Delta(z)}
	=\frac{\Delta_{\lambda+\delta}(z)}{\Delta_\delta(z)}
	=F_\lambda\left(
		-\sum_1^N z_i,-\frac{1}{2}\sum_1^N z_i^2,
		-\frac{1}{3}\sum_1^N z_i^3,\ldots
	\right),
$$
establishing the first equality of Lemma~\ref{L2.5}. In order to establish
the second one, note
\begin{eqnarray*}
	\tr\left(\frac{(Y+Z)^{p+1}}{p+1}\right)_2
	&=&\frac{1}{2}\tr(Y^2Z^{p-1}+YZYZ^{p-2}+\cdots+YZ^{p-1}Y)\\
	&=&\frac{1}{2}\sum_{i,j}Y_{ij}Y_{ji}(z_i^{p-1}+z_i^{p-2}z_j
	+\cdots+z_j^{p-1})\\
	&=&\frac{1}{2}\sum_{i,j}Y_{ij}Y_{ji}
	\left(\frac{z_i^p-z_j^p}{z_i-z_j}\right).
\end{eqnarray*}
Hence, performing a Gaussian integration, we find
\begin{eqnarray*}
\int dY\exp\tr\left(-\frac{(Y+Z)^{p+1}}{p+1}\right)_2
&=&	\int dY\exp\biggl(
	-\frac{1}{2}\sum_{i,j}Y_{ij}Y_{ji}\frac{z_i^p-z_j^p}{z_i-z_j}
	\biggr)\\
&=&	(2\pi)^{N^2/2}\Biggl(\prod_{1\le i,j\le
	N}\frac{z_i-z_j}{z_i^p-z_j^p}\Biggr)^{1/2} \\
&=&	\frac{(2\pi)^{N^2/2}}{p^{N/2}}\prod_{1\le i<j\le
	N}\frac{z_i-z_j}{z_i^p-z_j^p}\prod_1^p z_i^{-\frac{p-1}{2}}\\
&=&	\frac{(2\pi)^{N^2/2}}{p^{N/2}}
	\frac{\Delta(z)}{\Delta(z^p)}\prod_1^N z_i^{-\frac{p-1}{2}},
\end{eqnarray*}
establishing Lemma~\ref{L2.5}.
\endproof

\begin{remark}
In general we have
$$
	\int_{\cal H}dY\,\E^{-\tr(V(Y+Z))_2}
	=(2\pi)^{N^2/2}\frac{\Delta(z)}{\Delta\bigl(V'(z)\bigr)}
	\frac{1}{\sqrt{\prod_1^NV''(z_i)}}\,.
$$
\end{remark}

The following lemma is due to Harish Chandra, Bessis--Itzykson--Zuber
and Duistermaat--Heckman among others:
\begin{lemma}\label{L2.6}
Given $N\times N$-diagonal matrices $X$ and $Y$, we have
$$
	\int_{{\bf U}(N)} \E^{\tr XUYU^\dagger} dU=(2\pi)^{\frac{N(N-1)}{2}}
	\frac{\det(\E^{x_iy_j})_{1\le i,j\le N}}{\Delta(X)\Delta(Y)}\,.
$$
\end{lemma}
A proof can be found in \cite{hc}.

\section{Matrix Fourier Transforms}\label{Kontsevich}

In this section we explain how generalized Kontsevich integrals (see
\cite{kon,avm,AM}) are closely related to the theory of Fourier transforms.
Indeed, if $V(x)$ grows sufficiently at infinity, any {\em Fourier transform\/}
\begin{equation}
	a(y)=\int_{-\infty}^\infty \E^{-V(x)+xy}dx\label{3.1}
\end{equation}
leads to a linear space of functions ${\cal W}$ {\em invariant under two
operators $A$ and\/} $V'(z)$ satisfying $[A,V'(z)]=1$.
\medbreak\noindent
(i) The point is that $a(y)$ satisfies the differential equation
\begin{equation}
	V'\left(\frac{\partial}{\partial y}\right)a(y)=ya(y)\,,
	\label{3.2}
\end{equation}
as seen from
\begin{eqnarray*}
0&=&	\int_{-\infty}^\infty\frac{\partial}{\partial x}\E^{-V(x)+xy}dx
	=\int_{-\infty}^\infty\bigl(-V'(x)+y\bigr)\E^{-V(x)+xy}dx\\
&=&	\left(-V'\left(\frac{\partial}{\partial y}\right)+y\right)a(y).
\end{eqnarray*}
Thus setting $y=V'(z)$ in (\ref{3.2}) and
$A_0:=V''(z)^{-1}\partial/\partial z = \partial/\partial y\bigr|_{y=V'(z)}$,
the function $a\bigl(V'(z)\bigr)$ satisfies the
differential equation
\begin{equation}
	V'(A_0)a\bigl(V'(z)\bigr)=V'(z)a\bigl(V'(z)\bigr)\,.\label{3.3}
\end{equation}
(ii) The method of stationary phase applied to integrals (\ref{3.1})
and their derivatives leads to the following estimate,
upon Taylor expanding $V(x)$ around $x=z$,
\begin{eqnarray}
\lefteqn{
	\left(\frac{\partial}{\partial y}\right)^na(y)\biggr|_{y=V'(z)}
}\nonumber\\
&=&	\int_{-\infty}^\infty x^n\E^{-V(x)+xV'(z)}dx
	\nonumber\\
&=&	\int_{-\infty}^\infty x^n
	\E^{-\left(V(z)+(x-z)V'(z)+(1/2)(x-z)^2V''(z)+O(x-z)^3\right)+xV'(z)}
	dx
	\nonumber\\
&=&	\E^{-V(z)+zV'(z)}\int_{-\infty}^\infty x^n\E^{-(1/2)(x-z)^2V''(z)
	\left(1+(V'''/V'')O(x-z)\right)}dx
	\nonumber\\
&=&	\E^{-V(z)+zV'(z)}\frac{1}{\sqrt{V''}}\left(
	\int_{-\infty}^\infty\left(\frac{y}{\sqrt{V''}}+z\right)^n
	\E^{-y^2/2}dy+O(1/z)\right)
	\nonumber\\
&=&	\rho(z)^{-1}z^n\bigl(1+O(1/z)\bigr)\,,
	\label{3.4}
\end{eqnarray}
with
$$
	\rho(z)=\frac{1}{\sqrt{2\pi}}\E^{V(z)-zV'(z)}\sqrt{V''(z)}\;.
$$
Therefore defining
$$
	A:=\rho(z)\frac{\partial}{\partial y}\biggr|_{y=V'(z)}
	\circ\rho(z)^{-1}
$$
and
$$
\psi_n(z):=A^n\psi_0(z):=\rho(z)\frac{\partial^n}{\partial y^n}
a(y)\biggr|_{y=V'(z)}\,,\quad n=0,1,\ldots,
$$
the differential equation (\ref{3.3}) implies
$$
	V'(A)\psi_0(z)=V'(z)\psi_0(z)\,.
$$
This, combined with (\ref{3.4}), proves that the linear span
$$
	{\cal W}:=\Span_\BC\bigl\{\psi_k(z)=z^k\bigl(1+O(1/z)\bigr)\bigm|
	k=0,1,2,\ldots\bigr\}
$$
is invariant under the operators $A$ and $V'(z)$, i.e.,
$$
	A{\cal W}\subset
	{\cal W}\quad\mbox{and}\quad
	V'(z){\cal W}\subset {\cal W}\,,\quad\mbox{with}\quad[A,V'(z)]=1\,.
$$
(iii) By Lemma~\ref{L2.4}, the $\tau$-function
corresponding to ${\cal W}$, at time $t$ as in (\ref{miwa}), is given by
\begin{eqnarray*}
\tau(t)&=&\frac{
		\det\bigl(A^{j-1}\psi_0(z_i)\bigr)_{1\le i,j\le N}
	}{
		\det(z_i^{j-1})_{1\le i,j\le N}
	}\\
&=&	\frac{1}{\Delta(z)}\det\left(
		\rho(z_i)\biggl(\frac{\partial}{\partial y}\biggr)^{j-1}
		\int_{-\infty}^\infty \E^{-V(x)+xy}dx\biggr|_{y=V'(z_i)}
	\right)_{1\le i,j\le N}\\
&=&	\frac{\prod^N_1\rho(z_i)}{\Delta(z)}
	\int_{\BR^N}dx\,\E^{-\sum_1^NV(x_i)}
	\Delta(x)\prod^N_1\E^{x_\alpha V'(z_\alpha)}\\
&=&	\frac{\prod^N_1\rho(z_i)}{N!\Delta(z)}
	\int_{\BR^N}dx\,\E^{-\sum_1^NV(x_i)}\Delta(x)
	\det\bigl(\E^{x_\alpha V'(z_\beta)}\bigr)_{1\le\alpha,\beta\le N}\,,\\
& &	\qquad\mbox{using Lemma~\ref{L2.3} with $s=1$ and the
	skew-symmetry of $\Delta(x)$},\\
&=&	\frac{\prod_1^N\rho(z_i)}{N!\Delta(z)/\Delta\bigl(V'(z)\bigr)}
	\int_{\BR^N}dx\,\E^{-\sum_1^NV(x_i)}\Delta^2(x)
	\frac{
		\det\bigl(\E^{x_\alpha V'(z_\beta)}
		\bigr)_{1\le\alpha,\beta\le N}
	}{
		\Delta(x)\Delta\bigl(V'(z)\bigr)
	}\\
&=&	c\frac{\prod_1^N\rho(z_i)}{\Delta(z)/\Delta\bigl(V'(z)\bigr)}
	\int_{\BR^N}dx\,\E^{-\sum_1^NV(x_i)}\Delta^2(x)
	\int_{{\bf U}(N)}dU\,\E^{\tr UXU^{-1}V'(Z)}\,,\\
& &	\qquad\mbox{using Lemma~\ref{L2.6}, with $X=\diag(x)$},\\
&=&	c'\E^{\tr(V(Z)-ZV'(Z))}
	\frac{
		\int_{{\cal H}_N}dX\,\E^{-\tr V(X)}\E^{\tr XV'(Z)}
	}{
		\int_{{\cal H}_N}dX\,\E^{-\tr(V(X+Z))_2}
	}\,,
	\quad\mbox{using Lemma~\ref{L2.5}},\\
&=&	c''\frac{
		\int_{{\cal H}_N} dY\,\E^{-\tr(V(Y+Z))_{\ge 2}}
	}{
		\int_{{\cal H}_N} dY\,\E^{-\tr(V(Y+Z))_2}
	}\,,\quad\mbox{upon setting}\ X=Y+Z,
\end{eqnarray*}
for some constants $c$, $c'$ and $c''$ depending on $N$.

\section{Generalized H\"ankel Functions, Differential Equations and
Laplace Transforms}
\label{sect4}

This section deals with the properties of H\"ankel functions and
their generalizations.

\begin{figure}
\centerline{\psfig{figure=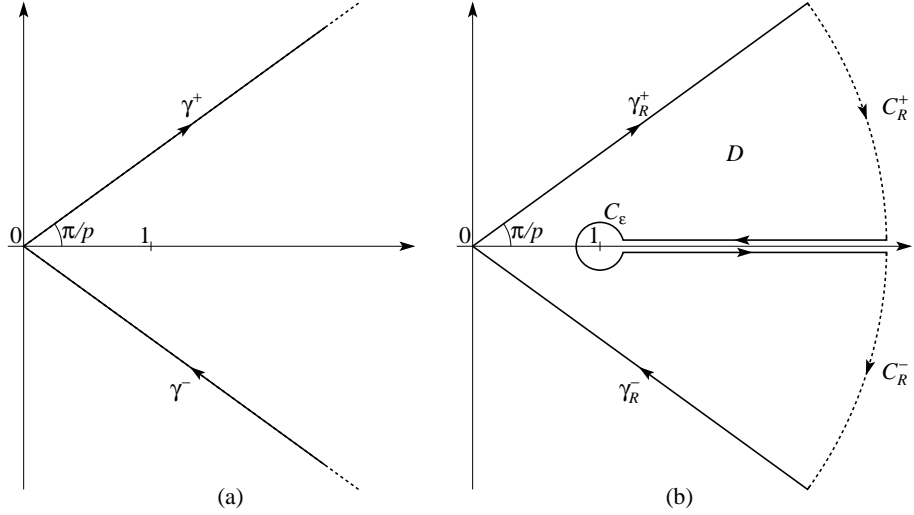}}	% to use Darrell's psfig macros
\caption{Contours of integration: (a) contour $\gamma$ for
(\protect{\ref{4.1.5}}); (b) closed contour for (\protect{\ref{cauchy}})}
\label{F1}
\end{figure}

\begin{lemma}\label{L4.1} The family of integrals
\begin{equation}
	\psi_k(z)=\frac{p^{c+1}}{\Gamma(-c)}\int_1^\infty
	\frac{z^{-c}(uz)^k\E^{-(u-1)z}}{(u^p-1)^{c+1}}\,du\,,\ \
	\begin{array}[t]{l}
	-1<c<0,\\ k=0,1,\dots,\ p=2,3,\ldots
	\end{array}
	\label{4.1}
\end{equation}
admits, for large $z>0$, an asymptotic expansion in $\BC((z^{-1}))$
of the form
\begin{equation}
	\psi_k(z)=z^k\bigl(1+O(1/z)\bigr)\,,
	\label{4.2}
\end{equation}
with $\psi_0(z)$ satisfying the differential equation
$$
	\E^zz^{-c}\left(\prod_{i=0}^{p-1}\left(z\frac{\partial}{\partial
	z}-i\right)-cp\prod_{i=0}^{p-2}
	\left(z\frac{\partial}{\partial z}-i\right)
	\right)z^c \E^{-z}\psi_0(z)=(-z)^p\psi_0(z)\,,
	\eqno(\ref{1.3})
$$
or equivalently
$$
	\E^zz^{-c}\left(
		z^p\left(\frac{\partial}{\partial z}\right)^p-cp\,z^{p-1}
		\left(\frac{\partial}{\partial z}\right)^{p-1}
	\right)
	z^c \E^{-z}\psi_0(z)=(-z)^p\psi_0(z)\,.
	\eqno(\ref{1.3}')
$$
Moreover $\psi_k(z)$ admits the following representation in terms of a
double integral\/\footnote{
	If $p=2$, so that $\gamma$ becomes the imaginary axis, these
	integrals should be interpreted by replacing $\zeta$ by
	$\zeta_\varepsilon=\E^{(\pi \I/2)-\varepsilon}$, and $\gamma$ by
	$\BR_+\zeta_\varepsilon+\BR_+\zeta_\varepsilon^{-1}$, and then
	taking the limit as $\varepsilon\downarrow0$.}
\begin{eqnarray}
\psi_k(z)&=&\frac{p^{c+1}}{2\pi \I}z^{(p-1)(c+1)}
	\int_\gamma dw\int_0^\infty dx\,\E^{z-w} w^kx^c\E^{x(w^p-z^p)}
	\nonumber\\
&=&	\frac{p^{c+1}}{2\pi \I}z^{(p-1)(c+1)}\E^z\int_0^\infty dx\,x^c
	\E^{-xz^p}\int_0^\infty dy\,f_k(y)\E^{-xy^p}\,,
\label{4.1.5}
\end{eqnarray}
where, in the first integral, $\gamma :=\gamma^++\gamma^-\subset\BC$
denotes the contour consisting of two
half-lines $\gamma^\pm=\BR_+\zeta^{\pm 1}$, $\zeta:= \E^{\pi \I/p}$, through
the origin making an angle $\pm\pi/p$ with the positive real axis, with
the orientation given as to go from $\zeta^{-1}\cdot\infty$ to $0$ to
$\zeta\cdot\infty$ (see Fig.~\ref{F1}~(a)), and where in the second
integral,
$$
	f_k(y)=(\zeta^{k+1}\E^{-\zeta y} -\zeta^{-k-1}\E^{-\zeta^{-1}y})y^k
	=\sum^\infty_{j=0}\frac{(-1)^j}{j!}a_{j+k+1}y^{j+k},
$$
where $a_n=\zeta^n-\zeta^{-n}=2\I\sin(n\pi/p)$.
\end{lemma}

\proof
Setting $v=(u-1)z$, and using
$$
	\Gamma(-c)=\int_0^\infty\frac{\E^{-v}}{v^{c+1}}\,dv
	\quad\mbox{for}\ c<0,
$$
we first observe that for each $n\ge0$,
\begin{eqnarray*}
\psi_k(z)&=&
	\frac{p^{c+1}z^k}{\Gamma(-c)}\int_0^\infty
	\frac{
		(1+v/z)^k\E^{-v}
	}{
		v^{c+1}p^{c+1}\left(
			1+{1\over p}
			\left(\sum_{i=2}^p{p\choose i}(v/z)^{i-1}\right)
		\right)^{c+1}
	}\,dv\\
&=&	z^k\left(
		1+\tilde b_{k,1}z^{-1}+\cdots+\tilde b_{k,n}z^{-n}
		+O(1/z^{n+1})
	\right)
\end{eqnarray*}
as $z\to\infty$, where the
$\tilde b_{k,i}:=\bigl(\Gamma(-c+i)/\Gamma(-c)\bigr)b_{k,i}=
\left(\prod_{j=0}^{i-1}(-c+j)\right)b_{k,i}$ are obtained from the
coefficients $b_{k,i}$ of the expansion\footnote{
	Noting that the radius of convergence of this power series is
	$|\zeta-1|$, one can get a precise growth estimate of the
	coefficients of $\psi_k(z)$ which implies that, in particular,
	as always with the string equation, ${\cal W}$ does not belong to
	the $L^2$-Grassmannian of Segal--Wilson \cite{sw}.}
\begin{equation}
	\frac{(1+s)^k}{\left(
	1+{1\over p}\left(\sum_{i=2}^p{p\choose i}s^{i-1}\right)
	\right)^{c+1}}
	=1+\sum_{i=1}^\infty b_{k,i}s^i\,,
\label{ExpandedIntegrand}
\end{equation}
confirming the asymptotic expansion (\ref{4.2}).

Moreover, setting
$$
\varphi_0(z)=\int_1^\infty\frac{z^{-c}\E^{-uz}}{(u^p-1)^{c+1}}\,du\,,
$$
we have for $c<0$ and Re $z>0$,
\begin{eqnarray*}
0&=&	-z^{p-1-c}\frac{\E^{-uz}}{(u^p-1)^c}\biggr|_{u=1}^{u=\infty}\\
&=&	-z^{p-1}\int_1^\infty\frac{\partial}{\partial u}
	\left((u^p-1)\frac{z^{-c}\E^{-uz}}{(u^p-1)^{c+1}}\right)du\\
&=&	(-1)^p\int_1^\infty\left((-zu)^p-cp(-zu)^{p-1}-(-z)^p\right)
	\frac{z^{-c}\E^{-uz}}{(u^p-1)^{c+1}}\,du\\
&=&	(-1)^pz^{-c}\left(
		z^p\left(\frac{\partial}{\partial z}\right)^p
		-cp\,z^{p-1}\left(\frac{\partial}{\partial z}\right)^{p-1}
		-(-z)^p
	\right)z^c\varphi_0(z)\\
&=&	(-1)^pz^{-c}\left(
		\prod_{i=0}^{p-1}\left(z\frac{\partial}{\partial z}
		-i\right)-cp\prod_{i=0}^{p-2}\left(
			z\frac{\partial}{\partial z}-i
		\right)-(-z)^p
	\right)z^c\varphi_0(z)\,,
\end{eqnarray*}
using in the last line the operator identity
$$
	\prod_{i=0}^{p-1}\left(z\frac{\partial}{\partial z}-i\right)
	=z^p\left(\frac{\partial}{\partial z}\right)^p,
$$
thus showing that $\psi_0(z)$ satisfies the differential equation
(\ref{1.3}) or (\ref{1.3}$'$).

Consider a bounded domain $D\subset\BC$, whose boundary consists of
the lines $\gamma_R^\pm$, making an angle $\pm\pi /p$ with the
positive real axis, two circle segments $C_R^\pm$, about the
origin, of large enough radius $R$ and a small circle about 1 of
radius $\varepsilon$ connected to $C_R^\pm$, as in Fig.~\ref{F1}~(b).
The function
$\E^{-uz}/(u^p-1)^{c+1}$ is univalued in $D$ and all its
singularities lie outside $D$. By Cauchy's theorem we have
\begin{equation}
	\left(	\int_{\gamma_R^-}+\int_{\gamma_R^+}+\int_{C_R^+}+
		\int_R^{1+\varepsilon}+\int_{C_\varepsilon}+
		\int_{1+\varepsilon}^R+\int_{C^-_R}
	\right)
	\frac{(uz)^k\E^{-(u-1)z}}{(u^p-1)^c}\,du=0\,.
	\label{cauchy}
\end{equation}
Observe that, for $z>0$ and $p>2$, we have $z\cos\theta \ge z\cos(\pi/p) >0$
for $0\le\theta\le\pi/p$, implying
$$
	\int_{C_R^\pm} \frac{(uz)^k\E^{-(u-1)z}}{(u^p-1)^{c+1}}\,du
	=O\bigl(R^{k-(c+1)p+1}\E^{-Rz\cos(\pi/p)}\bigr) \to0
$$
as $R\uparrow\infty$.  Since $c<0$, we also have
$$
	\int_{C_\varepsilon}\frac{(uz)^k\E^{-(u-1)z}}{(u^p-1)^{c+1}}\,du
	=O(\varepsilon^{-c})\to0
$$
as $\varepsilon\downarrow0$.
So, taking limits as $\varepsilon\downarrow 0$ and $R\uparrow\infty$ leads
to
\begin{eqnarray*}
\int_\gamma\frac{(uz)^k\E^{-(u-1)z}}{(u^p-1)^{c+1}}\,du
&=&	-\left(\int_\infty^1 + \int_{1-\I0}^{\infty-\I0}\right)
	\frac{(uz)^k\E^{-(u-1)z}}{(u^p-1)^{c+1}}\,du\\
&=&	(1-\E^{-2\pi \I(c+1)})\int_1^\infty
	\frac{(uz)^k\E^{-(u-1)z}}{(u^p-1)^{c+1}}\,du\\
&=&	2\I\E^{-\pi \I c}\sin\pi c\int_1^\infty
	\frac{(uz)^k\E^{-(u-1)z}}{(u^p-1)^{c+1}}\,du\,.
\end{eqnarray*}
Note that, since $u^p-1<0$ along $\gamma$, we have the following
$\Gamma$-function representation
$$
	\frac{1}{(u^p-1)^{c+1}}=-\frac{\E^{-\pi \I c}}{\Gamma(c+1)}
	\int_0^\infty dx\,x^c\E^{x(u^p-1)}\,,
$$
and thus
\begin{eqnarray*}
\psi_k(z)&=&\frac{p^{c+1}}{\Gamma(-c)}\int_1^\infty\frac{z^{-c}(uz)^k
	\E^{-(u-1)z}}{(u^p-1)^{c+1}}\\
&=&	\frac{p^{c+1}\E^{\pi \I c}}{2\I \sin\pi c\,\Gamma(-c)}
	z^{-c}\int_\gamma
	\frac{(uz)^k \E^{-(u-1)z}}{(u^p-1)^{c+1}}\,du\\
&=&	-\frac{p^{c+1}z^{-c}}{2\I \sin\pi c\Gamma(-c)\Gamma(c+1)}
	\int_\gamma du\,(uz)^k\E^{-(u-1)z}
	\int_0^\infty dx\,x^c\E^{x(u^p-1)}\\
&=&	\frac{p^{c+1}}{2\pi \I}
	z^{-c}\int_\gamma du\,(uz)^k\E^{-(u-1)z}
	\int_0^\infty dx\,x^c \E^{x(u^p-1)}\,,\\
&=&	\frac{p^{c+1}}{2\pi \I}z^{(p-1)(c+1)}
	\int_\gamma dw\int_0^\infty dx\,\E^{z-w}w^kx^c\E^{xw^p}\E^{-xz^p}\,,
\end{eqnarray*}
upon setting $w=uz$. Here we used the $\Gamma$-function duplication,
$\Gamma(-c)\Gamma(c+1)=-\pi/\sin\pi c$, $-1<c<0$.
Working out the integral over $\gamma$, interchanging the
integrations and using $\zeta^{\pm p}=-1$, we find
\begin{eqnarray*}
\psi_k(z)&=&\frac{p^{c+1}}{2\pi \I}
	z^{(p-1)(c+1)}\E^z\int_0^\infty dx\,x^c \E^{-xz^p}\cdot\\
&&	\cdot\left(\zeta^{-k-1}\int_\infty^0dy\,\E^{-\zeta^{-1}y}y^k\E^{-xy^p}
	+\zeta^{k+1}\int_0^\infty dy\,\E^{-\zeta y}y^k\E^{-xy^p}\right)\\
&=&	\frac{p^{c+1}}{2\pi \I}z^{(p-1)(c+1)}\E^z\int_0^\infty dx\,x^c
	\E^{-xz^p}\int_0^\infty dy\,f_k(y)\E^{-xy^p}
\end{eqnarray*}
with
\begin{eqnarray*}
f_k(y)&=&(\zeta^{k+1}\E^{-\zeta y}-\zeta^{-k-1}\E^{-\zeta^{-1}y})y^k\\
&=&	\sum_{j=0}^\infty\frac{(-1)^j}{j!}(\zeta^{j+k+1}-\zeta^{-j-k-1})
	y^{j+k},
\end{eqnarray*}
as announced in (\ref{4.1.5}), thus ending the proof of Lemma~\ref{L4.1}.
\endproof
\begin{lemma}\label{L4.2} The linear space spanned by the generalized
H\"ankel functions,
$$
	{\cal W}=\Span_\BC\biggl\{\psi_k(z)
	=\frac{p^{c+1}}{\Gamma(-c)}\int_1^\infty
	\frac{z^{-c}(uz)^k\E^{-(u-1)z}}{(u^p-1)^{c+1}}\,du
	\biggm|k=0,1,2,\ldots\biggr\}
$$
is invariant under
$$
	z^p\quad\mbox{and}\quad
	A_c:= z^{-c}\E^zz\frac{\partial}{\partial z}\circ \E^{-z}z^c=
	z\frac{\partial}{\partial z}-z+c
$$
(so that $[(1/p)A,z^p]=z^p$),
with $\psi_0$ satisfying the differential equation (\ref{1.3}).
\end{lemma}

\proof
The space ${\cal W}$ is invariant under $A_c$,
because
\begin{eqnarray*}
A_c\psi_k(z)&=&\frac{p^{c+1}}{\Gamma(-c)}z^{-c}\E^z
	z\frac{\partial}{\partial z} z^c \E^{-z}\int_1^\infty
	\frac{z^{-c}(uz)^k\E^{-(u-1)z}}{(u^p-1)^{c+1}}\,du \\
&=&	\frac{p^{c+1}}{\Gamma(-c)}z^{-c}\E^z
	z\frac{\partial}{\partial z}\int_1^\infty
	\frac{(uz)^k\E^{-uz}}{(u^p-1)^{c+1}}\,du \\
&=&	k\psi_k(z)-\psi_{k+1}.\\
\end{eqnarray*}
Moreover, the operator
$$
	\prod^{p-1}_{i=0}(A_c-i)-cp\prod^{p-2}_{i=0}(A_c-i)
$$
has the form $\sum^p_0\alpha_jA_c^j$, with $\alpha_p=1$. From
Lemma~\ref{L4.1}, the
solution to the differential equation
$$
	\left(\prod^{p-1}_{i=0}(A_c-i)-cp\prod^{p-2}_{i=0}(A_c-i)\right)
	\psi_0(z)=(-z)^p\psi_0(z)
$$
is given by the function in (\ref{4.1}) or (\ref{4.1.5}) for $k=0$.
An asymptotic expansion of the form
$$
	\psi_0(z)=1+O(z^{-1})
$$
follows from (\ref{4.2}).
\endproof

\section{Proof of the Main Statements}

\subsection{Proof of Theorems~\protect{\ref{T2}} and \protect{\ref{T0}}
		and Remark~\protect{\ref{rmk1}}}

In Lemma~\ref{L4.2},
we have constructed a space ${\cal W}$ and an operator $A=A_c$ such
that
$$
	A{\cal W} \subset {\cal W}\quad\mbox{and}\quad
	z^p{\cal W} \subset {\cal W}\,,
$$
with the lowest order element $\psi_0\in{\cal W}$ satisfying
Eq.~(\ref{1.3}).
Proposition~\ref{stable2} and Remark~\ref{R3} imply
that the stabilizer of ${\cal W}$ is $\BC[A,z^p,z^{-p}F(A)]$, proving
Theorem~\ref{T2}, Part~(i).

Let $\Psi$ and $\tau$ be the wave function and the $\tau$-function,
respectively, associated with the KP time evolution
${\cal W}^t=e^{-\sum t_iz^i}\cal W$ of $\cal W$.
We now define the operators $P$ and $Q$ in the $x$-variable,
via the operators $A$ and $z^p$ in the $z$-variable, by means of
$$
	z^p\Psi(t,z)=P\Psi(t,z)\quad\hbox{and}\quad
	(1/p)A\Psi(t,z)=Q\Psi(t,z)\,.
$$
According to Lemma~\ref{L2.1}, $P$ and $Q$ are differential operators.
They satisfy $[P,Q]=P$ since $[(1/p)A,z^p]=z^p$. Note that $P$ and
$Q$ can also be written:
$$
	P=L^p=SD^pS^{-1}
$$
and
$$
	Q={1\over p}(ML-L+c)=
	\frac{1}{p}S\left(\sum_1^\infty k\bar t_kD^k-D+c\right)S^{-1},
$$
where
$$
	S=\frac{\tau(t-[D^{-1}])}{\tau(t)}
$$
in terms of the $\tau$-function above, and $L$ and $M$ are as in
(\ref{1.16}) and (\ref{1.27}), proving Theorem~\ref{T2}, Part~(ii).

Since $(M-1)L=pQ-c$ is a differential operator, we also have,
using the notation $\alpha_{ij}$ as in the statement of Theorem~\ref{T2},
\begin{eqnarray*}
	\bigl((M-1)L\bigr)^mL^{np}
	&=&\sum_{i=1}^m\alpha_{m,i}(M-1)^iL^{i+np}\\
	&=&\sum_{\scriptstyle0\le i \le m\atop\scriptstyle0\le j \le i}
	\alpha_{m,i}{i\choose j}(-1)^{i-j}M^jL^{i+np}
\end{eqnarray*}
is a differential operator.  Thus
$$
	\sum_{\scriptstyle 0 \le i \le m\atop\scriptstyle 0 \le j \le i}
	\alpha_{m,i}{i\choose j}(-1)^{i-j}(M^jL^{i+np})_-\Psi=0\,,
$$
implying (\ref{Wconstraint}), upon using (\ref{asvmformula}), completing
the proof of Theorem~\ref{T2}.

To prove Remark~\ref{rmk1}, we evaluate
$$
	\left(
	\sum it_i{\partial\over\partial t_i}-{\partial\over\partial t_1}-a
	\right)\tau=0
$$
at $t=0$ to find
$-\bigl((\partial\tau/\partial t_1)/\tau\bigr)\bigr|_{t=0}=a$.
Remember, on the one hand,
$$
	\Psi(0,0,z)=\psi_0(z)=(1+\tilde b_{01}z^{-1}+\cdots)\,,
$$
and on the other hand
\begin{eqnarray*}
	\Psi(0,0,z)
	&=&
	{\tau(t_1+x-z^{-1},\ldots)\over\tau(t_1+x,\ldots)}
	\biggr|_{x=0,t=0}\\
	&=& \biggl(
	1-\tau^{-1}{\partial\tau\over\partial x}z^{-1}+\cdots
	\biggr)\biggr|_{t=0}.
\end{eqnarray*}
Therefore $a=\tilde b_{01}=(-c)b_{0,1}=c(1+c)(p-1)/2$ as stated in
Remark~\ref{rmk1}, as implied by (\ref{ExpandedIntegrand}).

To prove Theorem~\ref{T0}, note that at $t=0$,
\begin{eqnarray*}
	Q|_{t=0}&=&(1/p)S((x-1)(\partial/\partial x)+c)S^{-1}\\
		&=&(1/p)(x-1)(\partial /\partial x)+c+
		(\mbox{negative order terms})\,.
\end{eqnarray*}
Since $Q$ must be a differential operator, the negative order terms
vanish, and $Q|_{t=0}=(1/p)(x-1)(\partial /\partial x) + c$. Thus, from
the second equation in (\ref{eigenProblem}), we have
\begin{equation}
	0=\bigl(Q|_{t=0}-A_c\bigr)\Psi(x,0,z)
	=\bigl((x-1)(\partial /\partial x)-z(\partial /\partial z-1))
	\Psi(x,0,z)\,.
\label{firstorder}
\end{equation}
Since this is a first order equation and the line $x=0$ is
noncharacteristic, $\Psi(x,0,z)$ is determined by (\ref{firstorder})
together with the initial condition $\Psi(0,0,z)=\psi_0(z)$.  It is
easy to check that the right-hand side of (\ref{similar}) satisfies
these conditions.  Finally, (\ref{1.3a}) follows from (\ref{1.3}):
Writing (\ref{1.3}) as
$$
	F\biggl(z{\partial\over\partial z}+c\biggr)\bigl(\E^{-z}\psi_0(z)\bigr)
	= (-z)^p\E^{-z}\psi_0(z)\,,
$$
substituting $(1-x)z$ for $z$, using the scaling invariance of
$z\,\partial/\partial z$, and dividing both sides by $z^p$, we get
\begin{equation}
	{1\over z^p}F\biggl(z{\partial\over\partial z} + c\biggr)
	\bigl(\E^{(x-1)z}\psi_0((1-x)z)\bigr)
	= (x-1)^p \E^{(x-1)z}\psi_0((1-x)z)\,.
\label{1.3x}
\end{equation}
Multiplying both sides of this formula by $\E^z$, and using the identity
$\E^z(z\,\partial/\partial z+c)=(z\,\partial/\partial z-z+c)\circ \E^z$,
we get the second formula in (\ref{1.3a}).
Next, switching the roles of $z$ and $1-x$ in (\ref{1.3x}), we get
$$
	{1\over(x-1)^p}F\biggl((x-1){\partial\over\partial x}+c\biggr)
	\bigl(\E^{(x-1)z}\psi_0((1-x)z)\bigr)
	= z^p \E^{(x-1)z}\psi_0((1-x)z)\,.
$$
Multiplying both sides of this formula by $\E^z$, and using the fact
that $\E^z$ commutes with $(x-1)\partial /\partial x + c$, we get
the first formula in (\ref{1.3a}), completing the proof of Theorem~\ref{T0}.

\subsection{Proof of Theorem~\protect{\ref{T1}}}

Setting $t_n=-\frac{1}{n}\sum_{i=1}^n z_i^{-n}$, $n=1$,~2, \dots, and
using Lemma~\ref{L2.4}, and Lemma~\ref{L2.3} with $s=2$, we have
\begin{eqnarray*}
\tau(t)
&=&	\frac{\det\bigl(\psi_{k-1}(z_i)\bigr)_{1\le k,i\le N}}{\Delta(z)}\\
&=&	\frac{a^N}{\Delta(z)} \det\left(z_i^{(p-1)(c+1)} \E^{z_i}
	\int_0^\infty dx
	\int_0^\infty dy\,x^c \E^{-xz_i^p}f_{k-1}(y)
	\E^{-x y^p}\right)_{k,i}\\
&=&	\frac{a^N S_2(t)}{\Delta(z)}
	\int_{\BR_+^N} dx \int_{\BR_+^N} dy \left(\prod_1^N x_i^c\right)
	\cdot\\
&&	\cdot
	\det\bigl(f_{k-1}(y_i)\bigr)_{k,i}\
	\E^{-\sum_1^Nx_iz_i^p} \E^{-\sum_1^N x_i y_i^p}\\
&=&	\frac{a^N S_2(t)}{(N!)^2\Delta(z)}
	\int_{\BR_+^N} dx \int_{\BR_+^N} dy \left(\prod_1^N x_i^c\right)
	\cdot\\
&&	\cdot
	\det\left(f_{k-1}(y_i)\right)_{k,i}
	\det\left(\E^{-x_iz_j^p}\right)_{i,j}
	\det\left(\E^{-x_i y_j^p}\right)_{i,j}\\
&=&	\frac{a^N S_2(t)\Delta(z^p)}{(N!)^2\Delta(z)}
	\int_{\BR_+^N} dx
	\int_{\BR_+^N} dy \left(\prod_1^N x_i^c\right)
	\Delta(x)^2 \Delta(y)^2
	\cdot\\
&&	\cdot
	S_0(y)\frac{
		\det\left(\E^{-x_i z_j^p}\right)_{i,j}
	}{	\Delta(x)\Delta(z^p) }
	\frac{
		\det\left(\E^{-x_i y_j^p}\right)_{i,j}
	}{	\Delta(x)\Delta(y^p) }\,,
\end{eqnarray*}
where $a=p^{c+1}/2\pi\I$,
$$
	S_2(t) = \prod_1^N\left(z_i^{(p-1)(c+1)}\E^{z_i}\right)\,,
$$
and
$$
	S_0(y_1,y_2,\ldots,y_N) =
	\frac{\Delta(y^p)}{\Delta(y)}
	\frac{\det\bigl(f_{k-1}(y_i)\bigr)_{1\le i,k\le N}}{\Delta(y)}\,.
$$
So we have, for some constants $C$, $C'$ and $C''$ depending on $N$, $p$ and $c$,
\begin{eqnarray*}
\tau(t) &=& C\frac{S_2(t)\Delta(z^p)}{\Delta(z)}
	\int_{\BR_+^N} dx\,\Delta(x)^2
	\int_{\BR_+^N} dy\,\Delta(y)^2 S_0(y)
	\cdot\\
&&	\cdot
	\int_{{\bf U}(N)} dU_X\,\E^{-\tr Z^pU_X^{-1} x U_X}
	\int_{{\bf U}(N)} dV_Y\,\E^{-\tr x V_Y^{-1} y^pV_Y}\\
&&	\qquad\mbox{using Lemma~\ref{L2.6}}\\
&=&	C\frac{S_2(t)\Delta(z^p)}{\Delta(z)}
	\int_{\BR_+^N} dx\,\Delta(x)^2 \left(\prod_1^N x_i^c\right)
	\int_{\BR_+^N} dy\,\Delta(y)^2 S_0(y)
	\cdot\\
&&	\cdot
	\int_{{\bf U}(N)} dU_X\,\E^{-\tr Z^pU_X^{-1} x U_X}
	\int_{{\bf U}(N)} dU_Y\,\E^{-\tr U_X^{-1} x U_X U_Y^{-1} y^p U_Y} \\
&&	\qquad\begin{tabular}{l}
		setting $U_Y = V_Y U_X$ for fixed $U_X$ in the last\\
		integral and noting that $dU_X\,dU_Y = dU_X\,dV_Y$
	\end{tabular}\\
&=&	C\frac{S_2(t)\Delta(z^p)}{\Delta(z)}
	\int_{\BR_+^N} dx\,\Delta(x)^2 \left(\prod_1^N x_i^c\right)
	\int_{{\bf U}(N)} dU_X\,\E^{-\tr Z^pU_X^{-1} x U_X}
	\cdot\\
&&	\cdot
	\int_{\BR_+^N} dy\,\Delta^2(y) S_0(y)
	\int_{{\bf U}(N)} dU_Y\,\E^{-\tr U_X^{-1} x U_X U_Y^{-1} y^p U_Y}\\
&=&	C'\frac{S_2(t)\Delta(z^p)}{\Delta(z)}
	\int_{{\cal H}_N^+} dX\,\det(X^c) \E^{-\tr Z^p X}
	\int_{{\cal H}_N^+} dY\,S_0(y)\E^{-\tr XY^p} \\
&=&	C''S_1(t)
	\frac{	\int_{{\cal H}_N^+} dX \det(X^c) \E^{-\tr Z^p X}
		\int_{{\cal H}_N^+} dY S_0(y) \E^{-\tr X Y^p}
	}{	\int_{{\cal H}_N} dX
		\exp\tr\left(-\frac{\left((X+Z)^{p+1}\right)_2}{p+1}\right)
	}\,,
\end{eqnarray*}
where we used Lemma~\ref{L2.5} in the last equality, and the definition
of $S_1(t)$ in Theorem~\ref{T1}.
A similar calculation, outlined below, implies the second formula for
$\tau$, upon using
the first representation of $\psi_k(z)$ in (\ref{4.1.5}):
\begin{eqnarray*}
	\tau(t)
	&=&\frac{\det\left(
		A^{k-1}\Psi(0,z_i)
	\right)_{1\le k,i\le N}}{\Delta(z)}\,,
	\quad\mbox{with $t_n=-\frac{1}{n}\sum^\infty_{i=1}z_i^{-n}$,}\\
	&=&\frac{1}{\Delta(z)}\det\left(
		a\,\E^{z_i}z_i^{(p-1)(c+1)}\int_\gamma dw
		\int_0^\infty dx\,\E^{-w}w^{k-1}x^c
		\E^{xw^p}\E^{-xz_i^p}
	\right)_{k,i}\\
	&=&\frac{a^N}{\Delta(z)}\E^{\sum z_i}\prod z_i^{(p-1)(c+1)}\cdot\\
	&&\cdot\int_\gamma\cdots\int_\gamma dw
	\int_0^\infty\cdots\int_0^\infty dx\,
	\E^{-\sum w_i}\prod
	x^c_i\Delta(w)\prod^N_{i=1}\E^{-z_i^px_i}\prod^N_{i=1}\E^{x_iw_i^p}\\
	&=&\frac{a^N}{(N!)^2}\E^{\sum z_i}\prod z_i^{(p-1)(c+1)}
	\frac{1}{\Delta(z)}\int_{\gamma^N}dw\int_{\BR_+^N}dx\,\E^{-\sum w_i}
	\prod x_i^c\Delta(w)\cdot\\
	&&\cdot
	\det\bigl(\E^{-z_i^px_j}\bigr)_{1\le i,j\le N}
	\det\bigl(\E^{x_i w_j^p}\bigr)_{1\le i,j\le N}\\
	&=&\cdots\\
	&=&\frac{
		\int_{{\cal H}^\gamma_N}m(dW) \int_{{\cal H}^+_N}dX
		\det X^c
		\bigl(\Delta(w^p)/\Delta(w)\bigr)
		\E^{\tr(Z-W)}\E^{\tr X(W^p-Z^p)}
	}{	\int_{{\cal H}_N}dX
		\exp\tr\left(-\frac{(X+Z)^{p+1}}{p+1}\right)_2
	}\,,
\end{eqnarray*}
ending the proof of Theorem~\ref{T1}.

\end{document}